\documentclass{scrartcl}

\KOMAoptions{paper=a4,paper=portrait,fontsize=10pt}
\usepackage[english]{babel}
\usepackage[ansinew]{inputenc}
\usepackage[T1]{fontenc}
\usepackage{helvet}
\usepackage[a4paper,portrait]{geometry}
\let\oldtwocolumn\twocolumn

\newif\iftwocolumn\twocolumntrue
\def\onecolumn{\twocolumnfalse}
\def\twocolumn{\twocolumntrue}
\usepackage{indentfirst}
\newlength{\OneColumnWidth}
\newlength{\TwoColumnWidth}
\makeatletter
\renewcommand\section{\scr@startsection{section}{1}{\z@}{-3.5ex \@plus -1ex \@minus -.2ex}{2.3ex \@plus.2ex}{\normalfont\bfseries}}
\renewcommand\subsection{\scr@startsection{subsection}{2}{\z@}{-3.5ex \@plus -1ex \@minus -.2ex}{2.3ex \@plus.2ex}{\normalfont\bfseries}}
\makeatother

\usepackage[
	colorlinks=true,
	urlcolor=blue, 
	filecolor=blue, 
	linkcolor=blue, 
	citecolor=blue, 
]{hyperref}
\usepackage{cite} 

\usepackage[singlelinecheck=true,font=small,labelfont=bf,format=plain]{caption} 
\usepackage[singlelinecheck=true,font=small,labelfont=bf,format=plain]{subfig}

\let\oldhline\hline
\renewcommand{\hline}{\oldhline\rule{0pt}{12pt}}
\usepackage{enumitem}
\setlist{nosep}
\setenumerate[1]{label=(\arabic*)}
\setenumerate[2]{label=(\Alph*)}

\usepackage{ifthen}
\usepackage{forloop}
\usepackage{etoolbox}
\usepackage{graphicx}
\usepackage[fleqn]{amsmath}

\renewcommand{\title}[1]{\def\inserttitle{#1}}
\newcommand{\email}[1]{\def\insertemail{#1}}
\renewcommand{\abstract}[1]{\def\insertabstract{#1}}
\def\insertjournal{}
\def\insertdoi{}
\def\insertarxiv{}
\newcommand{\journal}[3][accepted]{
	\def\tmpa{#1}\def\tmpb{submitted}\ifx\tmpa\tmpb\def\journalpre{Submitted to: }\else\def\tmpb{accepted}\ifx\tmpa\tmpb\def\journalpre{Accepted for publication in: }\else\def\tmpb{prepared}\ifx\tmpa\tmpb\def\journalpre{Prepared for submission to: }\else\def\journalpre{}\fi\fi\fi
	\if\relax\detokenize{#3}\relax\def\insertjournal{\journalpre#2}\else\def\insertjournal{\journalpre\href{#3}{#2}}\fi
}
\newcommand{\doi}[1]{\if\relax\detokenize{#1}\relax\def\insertdoi{}\else\def\insertdoi{DOI: \href{http://dx.doi.org/#1}{#1}}\fi}
\newcommand{\arxiv}[2]{\if\relax\detokenize{#2}\relax\def\insertarxiv{}\else\def\insertarxiv{arXiv: \href{https://arxiv.org/abs/#1}{#1 [#2]}}\fi}
\newcounter{authors}
\newcounter{addresses}
\newcounter{keywords}
\newcommand{\addauthor}[2]{\csdef{author\arabic{authors}}{#1}\csdef{authoraddress\arabic{authors}}{#2}\stepcounter{authors}}
\newcommand{\addaddress}[1]{\csdef{address\arabic{addresses}}{#1}\stepcounter{addresses}}
\newcommand{\addkeyword}[1]{\csdef{keyword\arabic{keywords}}{#1}\stepcounter{keywords}}
\newcommand{\maketitlesub}{
	\newcounter{i}
	\newcounter{j}
	\noindent\textbf{%
		\Large\inserttitle\\[1em]
		\large\csuse{author0}$^{\csuse{authoraddress0}}$%
		\forloop{i}{1}{\value{i} < \value{authors}}{%
			, \csuse{author\arabic{i}}$^{\csuse{authoraddress\arabic{i}}}$%
		}
	}\\[1em]
	\normalsize
	\setcounter{j}{0}
	\forloop{i}{0}{\value{i} < \value{addresses}}{%
		\stepcounter{j}
		\ifnum\value{addresses}>1$^{\arabic{j}}$\fi\,\csuse{address\arabic{i}}
		\ifthenelse{\value{j}<\value{addresses}}{\\}{}
	}
	\ifx\insertemail\empty\\[1em]\else\\[0.5em]E-mail address: \insertemail\\[1em]\fi
	\textbf{Abstract:} \insertabstract
	\ifthenelse{\value{keywords}=0}{}{
		\\[1em]
		Keywords: \csuse{keyword0}%
			\forloop{i}{1}{\value{i} < \value{keywords}}{%
				; \csuse{keyword\arabic{i}}%
			}
	}
}
\renewcommand{\maketitle}{\iftwocolumn\oldtwocolumn[\maketitlesub\vspace{1.5em}]\else\maketitlesub\fi}

\usepackage[headsepline]{scrlayer-scrpage}
\clearpairofpagestyles
\ihead{%
	\ifx\insertarxiv\empty%
			\ifx\insertjournal\empty\else\textnormal\insertjournal\fi%
	\else%
		\ifx\insertdoi\empty%
			\ifx\insertjournal\empty\else\textnormal\insertjournal\fi%
		\else%
			\ifx\insertjournal\empty\else\textnormal\insertjournal\fi%
			\ifx\textnormal\empty\else\linebreak\textnormal\insertdoi\fi%
		\fi%
	\fi%
}
\ohead{%
	\ifx\insertarxiv\empty%
		\ifx\textnormal\empty\else\textnormal\insertdoi\fi%
	\else%
		\ifx\insertdoi\empty%
			\ifx\insertarxiv\empty\else\textnormal\insertarxiv\fi%
		\else%
			\ifx\insertarxiv\empty\else\linebreak\textnormal\insertarxiv\fi%
		\fi%
	\fi%
}
\ifoot{}
\ofoot{}

\usepackage[hang]{footmisc}
\let\oldthebibliography\thebibliography

\renewcommand\thebibliography[1]{
	\small
	\oldthebibliography{#1}
	\setlength{\parskip}{0pt}
	\setlength{\itemsep}{0pt plus 0.3ex}
}
\newcommand{\bstindent}{99}

\newcommand{\bstauthor}{}

\newcommand{\bstjournal}{}

\newcommand{\bsttitle}{}
\newcommand{\bstvolume}{}
\newcommand{\bstyear}{}
\newcommand{\bbland}{and}

\usepackage{dblfloatfix} 
\allowdisplaybreaks 
\newcommand\fig[1]{Fig.~\ref{fig:#1}}
\newcommand\Fig[1]{Fig.~\ref{fig:#1}}

\renewcommand\sec[1]{Sec.~\ref{sec:#1}}
\newcommand\Sec[1]{Sec.~\ref{sec:#1}}
\newcommand\eqn[1]{Eq.~(\ref{eqn:#1})}
\newcommand\Eqn[1]{Eq.~(\ref{eqn:#1})}

\onecolumn 

\title{Do equidistant energy levels necessitate a harmonic potential?}

\addauthor{Fabian Teichert}{}
\addauthor{Eduard Kuhn}{}
\addauthor{Angela Thr\"anhardt}{}

\addaddress{Institute of Physics, Technische Universit\"at Chemnitz, 09107 Chemnitz, Germany}

\email{fabian.teichert@physik.tu-chemnitz.de}

\abstract{
Experimental results from literature show equidistant energy levels in thin Bi films on surfaces, suggesting a harmonic oscillator description.
Yet this conclusion is by no means imperative, especially considering that any measurement only yields energy levels in a finite range and with a nonzero uncertainty.
Within this study we review isospectral potentials from the literature and investigate the applicability of the harmonic oscillator hypothesis to recent measurements.
First, we describe experimental results from literature by a harmonic oscillator model, obtaining a realistic size and depth of the resulting quantum well.
Second, we use the shift-operator approach to calculate anharmonic non-polynomial potentials producing (partly) equidistant spectra.
We discuss different potential types and interpret the possible modeling applications.
Finally, by applying $n$th order perturbation theory we show that \textbf{exactly} equidistant eigenenergies cannot be achieved by polynomial potentials, except by the harmonic oscillator potential.
In summary, we aim to give an overview over which conclusions may be drawn from the experimental determination of energy levels and which may not.
}

\addkeyword{equidistant spectrum}
\addkeyword{anharmonic oscillator}
\addkeyword{shift operator}
\addkeyword{quantum well state}
\addkeyword{semiconductor physics}
\addkeyword{angle-resolved photoemission spectroscopy (ARPES)}

\journal[accepted]{Optical and Quantum Electronics}{} 
\arxiv{1910.12522}{quant-ph} 

\begin{document}

\maketitle

\section{Introduction}

Thin films on surfaces with thicknesses of some ten layers act as a quantum well state perpendicular to the film surface. The in-plane directions are bulk-like
and can be separated within the 3D Schr\"odinger equation. The variation over the different layers can then be described by a 1D model, most commonly by a square quantum well, where the eigenenergies increase quadratically and scale with $1/d^2$, $d$ being the well width. In many cases, however, a deviation from this standard
case is observed.
For epitaxially grown SnO$_2$, InSb, and Pb films with low charge densities, measurements with exponents about $1.5$ have been reported~\cite{SciRep.5.17424, PhysStatSolB.33.425, PhysRevB.66.233408}.
Furthermore, Kr\"oger et al.~\cite{PhysRevB.97.045403} and Hirahara et al.~\cite{PhysRevB.75.035422} investigated Bi films 
\cite{PhysRevLett.115.106803, PhysRevLett.93.046403, PhysRevLett.117.236402, JElectronSpec.201.98, NanoLett.12.1776}
and demonstrated that these exhibit equidistant energy levels with a $1/d$ thickness dependence
\cite{PhysRevB.97.045403, PhysRevB.75.035422}.
For this, Kr\"oger et al.\ prepared epitaxially grown Bi(111) films on Si(111) samples.
The thicknesses varied between 20 and 100 bilayers (i.e.\ $8\,\text{nm}$ to $40\,\text{nm}$).
The conductance $G$ was measured as a function of temperature $T$.
A general dependence was fitted to this data and an effective bandgap $E_\text{g}$ was extracted by describing the temperature dependence of the conductance by thermal excitation, $G\propto\text{exp}(E_\text{g}/k_\text{B}T)$, where $k_\text{B}$ is the Boltzmann constant.
Hirahara et al.\ prepared Bi(001) films grown on Si(111) samples.
The thicknesses varied from 7 to 40 bilayers (i.e.\ $2.8\,\text{nm}$ to $16\,\text{nm}$).
The energy levels were measured with angle-resolved photoemission spectroscopy (ARPES) and were found to be equidistant.
This motivates a description by the well-known harmonic oscillator potential in contrast to the usually found square well potential, which was also stated by Kr\"oger et al.
An explanation given by Hirahara et al.\ is the phase shift accumulation model, originating from the Bohr-Sommerfeld quantization condition, which yields the $1/d$-dependence if the dispersion near the Fermi energy is linear~\cite{PhysRevB.75.035422}.
Nevertheless, the question must be asked whether equidistant energy levels actually necessitate a harmonic potential or whether alternative potential shapes with equidistant levels exist, especially considering that only a limited energy range is examined and a finite measurement uncertainty is present.
The general answer is ``no''~\cite{CommMathPhys.82.471}, which was demonstrated by previous analytic derivations resulting in different potentials by generalizing the creation/annihilation operator.
For this purpose, the factorization method \cite{JMathPhys.25.3387, TheorChemAcc.110.403} or the more general shift-operator approach~\cite{Chaos.4.47, SovPhysJETP.75.446} was used.

First principle studies of any material are expected to result in a potential which oscillates with the atomic positions of the real structure which is put in.
As the harmonic oscillator is a valid model, an expectation could be that the potential of thin Bi films would on average and at a scale larger than the atomic structure be quadratic with additional spatial variations at smaller scale.
Within this manuscript we focus on the model point of view and calculate model potentials satisfying the previous expectation.
We hope that this publication will serve to interpret experimental results more carefully and clarify that even when the energy levels are known, a wide range of potentials must still be considered.
In the following, we assume equidistant energy levels and investigate what follows for the potential.
In \sec{Experiments} the experimental results are modeled by a truncated harmonic oscillator, yielding good agreement of the energy levels with the analytical harmonic oscillator case up to a certain energy and providing a more physical description including the specific layer thickness.
\Sec{Shift_operator} treats the 1D  Schr\"odinger equation as an inverse problem by calculating the potentials from the given equidistant energy spectrum.
This is done using the shift-operator approach.
We present some previous analytical results and some new numerical calculations and possible interpretations concerning real structures.
In \sec{Perturbation} we show by perturbation theory that polynomial potentials apart from the quadratic harmonic oscillator potential cannot exhibit an infinite number of \textbf{exactly} equidistant energy levels.

\section{Comparison with experiments}\label{sec:Experiments}

\begin{figure}[t]
	\centering
	\includegraphics[width=\OneColumnWidth]{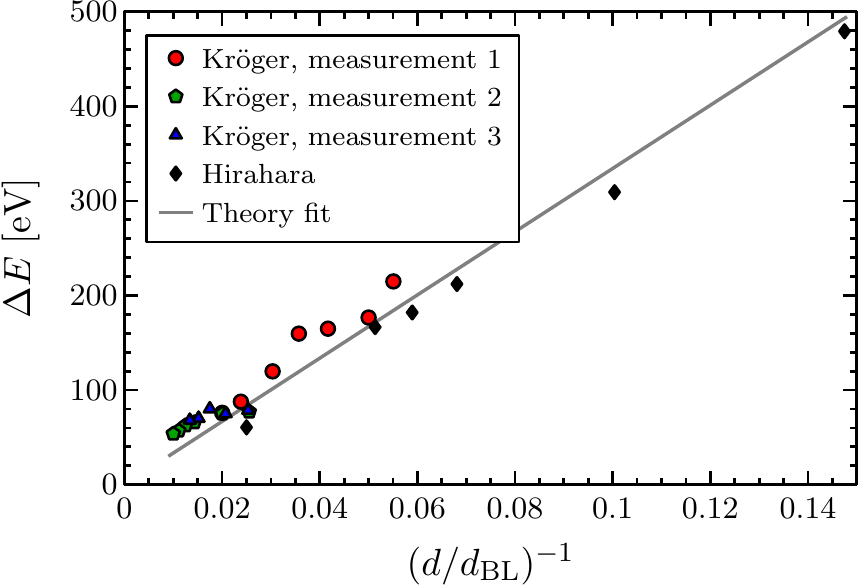}
	\caption{Experimental data of Kr\"oger et al.~\cite{PhysRevB.97.045403} (3 different Bi films) and Hirahara et al.~\cite{PhysRevB.75.035422} in comparison with a fit to the harmonic oscillator.}
	\label{fig:HO:experiment}
\end{figure}

\begin{figure}[b]
	\includegraphics[width=\TwoColumnWidth]{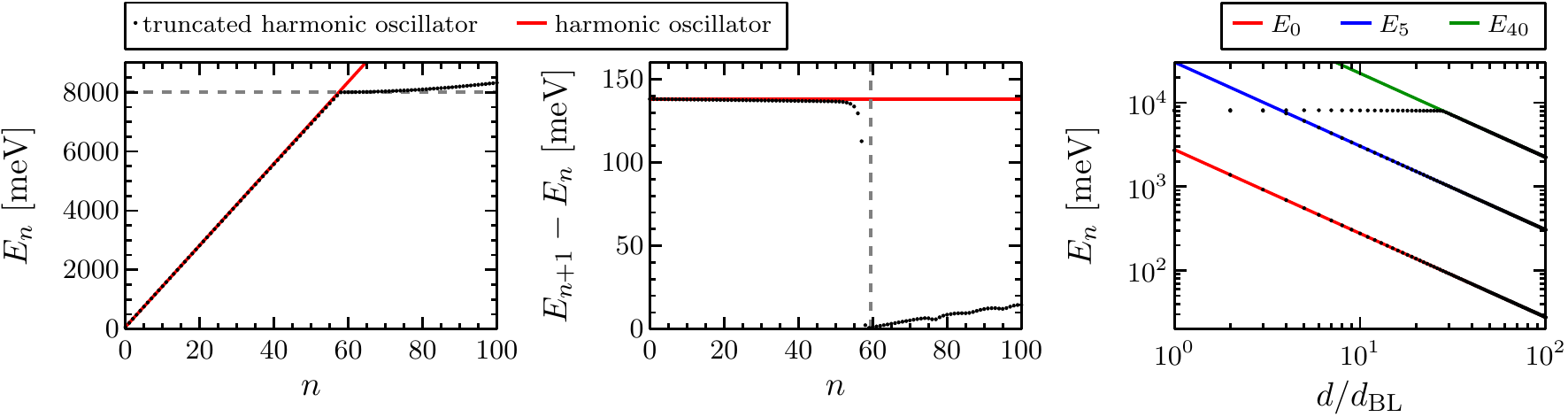}
	\caption{Left: energy levels. Center: energy level spacings. Right: thickness-dependent energy level. Colored lines denote the 1D harmonic oscillator. Black dots denote the truncated 1D harmonic oscillator of 16\,nm thickness.}
	\label{fig:HO:truncated}
\end{figure}

In this section, we concentrate on the interpretation of experimental results for Bismuth films from the literature.
The experimental results of Kr\"oger et al.~\cite{PhysRevB.97.045403} and Hirahara et al.~\cite{PhysRevB.75.035422}, i.e.\ the energy level spacings $\Delta E$ as a function of the Bi film thickness $d$, are depicted in \fig{HO:experiment} (colored symbols).
$d_\text{BL}$ is the thickness of a Bi bilayer, which form a stable unit in (111) direction.
For the Kr\"oger data, the measured bandgap $E_\text{g}$ is interpreted as energy spacing $\Delta E$.
The inverse thickness dependence can be clearly seen. The solid line shows a fit with a $d^{-1}$ thickness dependence.

To describe the experimental data and to connect the eigenenergies to a physical situation, the model of a truncated quadratic potential
\begin{flalign}
	V(x) &= V_0\begin{cases}
		1 & |x|\geq\frac{d}{2}\\
		\left(\frac{2x}{d}\right)^2 & |x|\leq\frac{d}{2}
	\end{cases} &
\end{flalign}
is used, where $d$ is the thickness of the material and the potential at the well edge $x=\pm\frac{d}{2}$ equals the well depth $V_0$.
Exemplary results are shown in \fig{HO:truncated} for $V_0/m^\ast=8\,\text{eV}/m_\text{e}$ and $d=16\,\text{nm}$, where $m^\ast$ is the effective mass and $m_\text{e}$ is the electron mass.
The energy eigenvalues $E_n$ and the energy level spacings $\Delta E$ of the truncated potential are in very good agreement with the non-truncated potential (i.e.\ the harmonic oscillator) for $V<0.9V_0$, yielding
\begin{align}
	E_n = \Delta E\left(n+\dfrac{1}{2}\right) \quad,\quad \Delta E = \frac{2\hbar}{d}\sqrt{\frac{2V_0}{m^\ast}} \label{eqn:fit}
\end{align}
with equidistant energy spacing $\Delta E$, which scales like $1/d$.

A regression according to~\eqn{fit} for reproducing the thickness dependence of the experimental data in \fig{HO:experiment} yields $\Delta E = 3.34\,\text{eV}\cdot d_\text{BL}/d$ and $V_0/m^\ast=2.93\,\text{eV}/m_\text{e}$.
All data are in good agreement with the regression. We conclude that the experimental data may be well fitted with a truncated harmonic oscillator potential which yields the desired equidistant energy states. Nevertheless, the question remains whether this is the only possible shape for the potential. The next section thus deals with finding alternative potential shapes with the same energy spectrum.

\section{Anharmonic oscillators with equidistant energy levels}\label{sec:Shift_operator}

\begin{figure}[t]
	\centering
	\includegraphics[width=\OneColumnWidth]{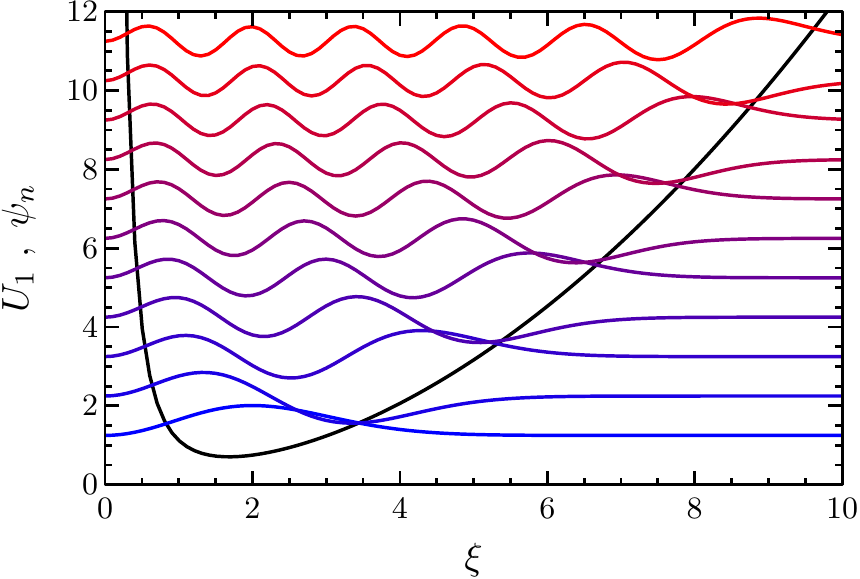}
	\caption{Potential and lowest states of the second-order shift operator for $A=1$. The offset of the states are the corresponding eigenenergies.}
	\label{fig:L2}
\end{figure}

In this section we show how the creation/annihilation operators of the harmonic oscillator can be generalized with the aim of finding further potentials with equidistant spectra.
An approach to model equidistant energy levels is using the shift operator to treat the Schr\"odinger equation as an inverse problem, explained in the following.
We first give a general introduction and show previous analytical results by Dubov et al.~\cite{Chaos.4.47, SovPhysJETP.75.446}.
Afterwards we present new numerical calculations and possible interpretations concerning real structures.

\subsection{Method}

The general idea is to define for an arbitrary potential the shift operator $\mathcal{L}$, which shifts the eigenstates $\psi$ in energy:
\begin{align}
	\mathcal{L}\psi(\xi,\epsilon) &= \psi(\xi,\epsilon+1) \quad.
\end{align}
Here, $\xi=\sqrt{m\omega/\hbar}x$ and $\epsilon=E/\hbar\omega$ are the dimensionless real space and energy.
If the states are equidistant, $\mathcal{L}$ transforms state $\psi_n$ into state $\psi_{n+1}$, resulting in the operator equation
\begin{align}
	[\mathcal{H},\mathcal{L}] &= \mathcal{L} \quad.
\end{align}
It can be solved by applying
\begin{align}
	\mathcal{L} = \sum_{k=0}^K \alpha_k(\xi)(\text{i}\mathcal{P})^k \quad \text{with}\quad K\geq 1 \quad.
\end{align}
$\mathcal{P}$ is the momentum operator.
$\alpha_k(\xi)$ are free parameters.

The first-order case ($K=1$) leads to the harmonic potential $U_0=\frac{1}{2}\xi^2$ and $\mathcal{L}$ equals the creation operator.
The derivations can be found in the Supplementary Material.
The solution of the second-order shift operator gives the isotonic oscillator~\cite{PhysLettA.70.177, IntJQuantumChem.110.1317, JPhysAMathGen.20.4331, JPhysAMathTheo.41.085301}
\begin{align}
	U_1 = \frac{1}{8}\xi^2 + \frac{A}{\xi^2} \label{eqn:U_2}
\end{align}
with either $\xi>0$ or $\xi<0$, depicted in \fig{L2} (black curve).
It can be interpreted as a half harmonic potential with a smooth wall, where the smoothness can be adjusted with $A$.
For the half harmonic oscillator (with a hard wall) only the odd solutions remain, resulting in a doubled energy level difference.
The potential in \eqn{U_2} yields very similar eigenstates and the same eigenenergies as the half harmonic oscillator, but slightly shifted in energy.
The eigenstates have a similar form to the harmonic potential:
\begin{align}
	\epsilon_n &= \frac{1}{2} + n + \frac{1}{4}\sqrt{1+8A} \quad,\\
	\psi_n(\xi) &= C_n \xi^{(1+\sqrt{1+8A})/2} \exp\left(-\frac{1}{4}\xi^2\right) \sum_{k=0}^n a_k \xi^{2k} \quad.
\end{align}
The eigenstates are also shown in \fig{L2}.
It can be concluded that with this potential it is possible to model asymmetric minima in contrast to the pure harmonic potential.

\begin{figure}[t]
	\centering
	\includegraphics[width=\TwoColumnWidth]{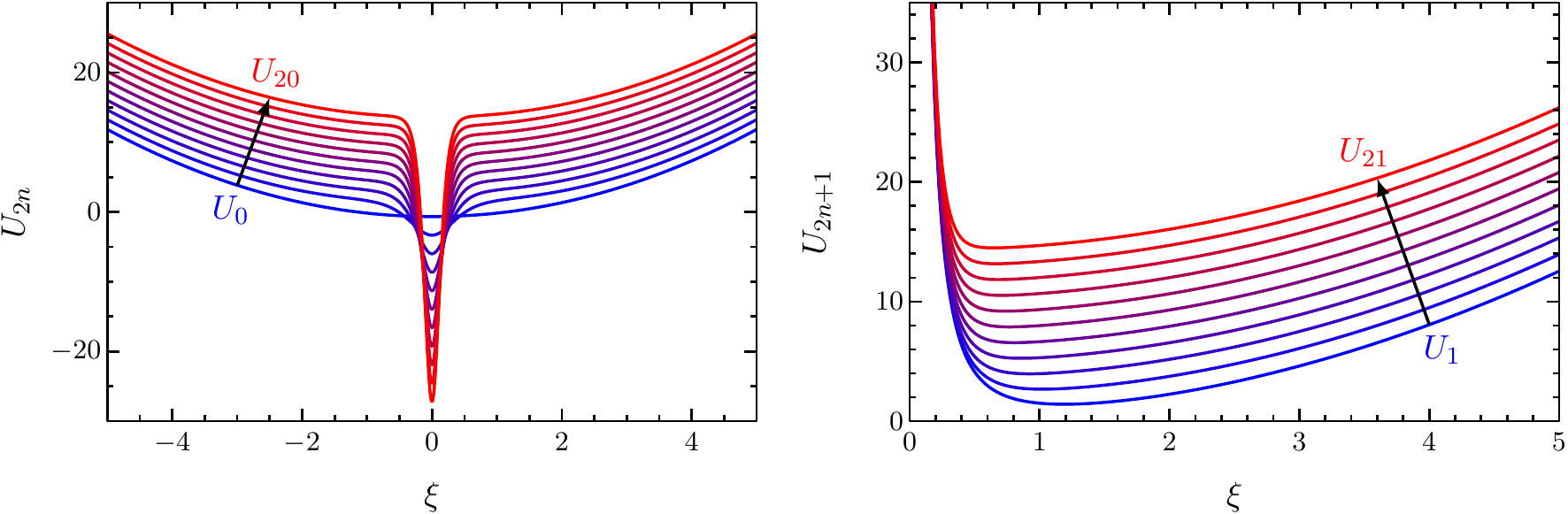}
	\caption{Potentials of the third-order shift operator $U_0 \ldots U_{21}$.}
	\label{fig:L3}
\end{figure}

The third-order shift operator results in the differential equation
\begin{gather}
	\frac{3}{2}\left(W^2\right)'' - \frac{1}{4}W'''' + \xi^2W'' + 3\xi W' = 0 \label{eqn:L3:U:DGL}
\end{gather}
with $U = W + \frac{1}{2}\xi^2$.
The prime denotes the first derivative.
 The equation can be solved as an initial value problem or a boundary value problem with any given conditions, resulting in many different solutions for the potential.
Possible solutions obtained by a Darboux transformation~\cite{Chaos.4.47, OsakaJMath.36.949} are
\begin{align}
	U_m &= -\frac{1}{2}\xi^2 - \frac{2}{3}(2m+1) + \left( \frac{P_m'(\xi)}{P_m(\xi)} + \xi \right)^2 \quad,\\
	P_{2m} &= \sum_{k=0}^m\frac{4^k}{(m-k)!(2k)!}\xi^{2k} \quad,\\
	P_{2m+1} &= \sum_{k=0}^m\frac{4^k}{(m-k)!(2k+1)!}\xi^{2k+1} \quad.
\end{align}
The index $m$ numbers the solutions.
They are drawn in \fig{L3}.
A further shift of the solutions in $\xi$ and $U$ is valid, but not a superposition due to the non-linearity of the constituting equations.
$U_{2m}$ are quadratic-like potentials with a dip at the minimum.
$U_0$ is the harmonic potential.
The corresponding energy levels are equidistant with $\Delta\epsilon=1$.
The ground state of potential $U_{2m}$ is additionally lowered by $2m$.
$U_{2m+1}$ are singular potentials at ${\xi=0}$.
$U_1$ is one solution of the second-order shift operator.
The corresponding energy levels are equidistant with $\Delta\epsilon=2$ ($\Delta\epsilon=1$ can be achieved by rescaling $\xi=\tilde{\xi}/2$).
The eigenstates of these anharmonic potentials are also similar to the ones of the harmonic potential.
Especially the potentials $U_1$ and $U_{2n+1}$ allow to construct asymmetric 1D models which describe equidistant energy levels, providing a much more flexible tool than a restriction to the analytical harmonic oscillator offers.

Further examples and their derivations can be found in the Supplementary Material.

\subsection{Numerical results}

\Eqn{L3:U:DGL} or its first integrals
\begin{gather}
	\frac{3}{2}\xi\left(W^2\right)' - \frac{3}{2}W^2 - \frac{1}{4}\xi W''' + \frac{1}{4}W'' + \xi^3W' = A \quad,\label{eqn:L3:U:DGL:2}\\
	-\frac{1}{2}\left( \frac{A + \frac{3}{2}W^2 - \frac{1}{4}W''}{\xi} \right)^2 - \frac{1}{2}W^3 + \frac{1}{8}\left(W'\right)^2 = AW + B\label{eqn:L3:U:DGL:3}
\end{gather}
can also be solved numerically to achieve further types of potentials.
We calculated the potential using the explicit Runge--Kutta method of 4th order or higher as implemented in the software package Mathematica~\cite{Mathematica.12.1}.
We quantify the numerical error of the solution using the local relative residuum, i.e. the residuum normalized to the maximum absolute additive contribution to the differential equation, evaluated as a function of $\xi$.
We get values below $10^{-4}$, which is sufficiently small to consider the solution correct.
The Schr\"odinger equation is solved by discretizing the differential operator and the potential on an equidistant grid ($\Delta x<0.01\,\text{nm}$) and solving the resulting matrix eigenvalue problem as implemented in Mathematica.
For this, the calculated spatial dimension is much larger than presented in the following, i.e. large enough for the potential at the boundary to be higher than the highest considered energy level, to prevent numerical errors from limiting the spatial dimension.

\begin{figure}[t]
	\includegraphics[width=\TwoColumnWidth]{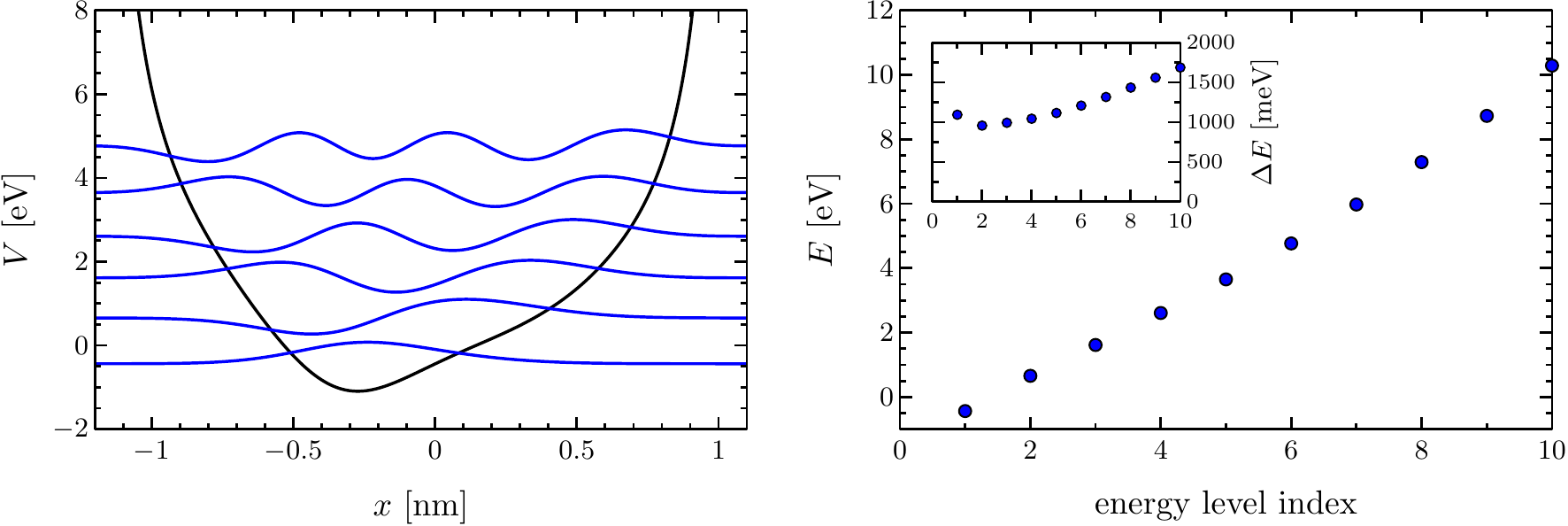}
	\caption{Numerical solution of \eqn{L3:U:DGL:2} with $A=-0.4$ and $W(1)=W'(1)=W''(1)=0$. Left: type-1 potential and lowest states. The offset of the states are the corresponding eigenenergies. Right: energy levels. The inset shows the energy level spacings.}
	\label{fig:L3:num_1}
\end{figure}

\begin{figure}[!b]
	\includegraphics[width=\TwoColumnWidth]{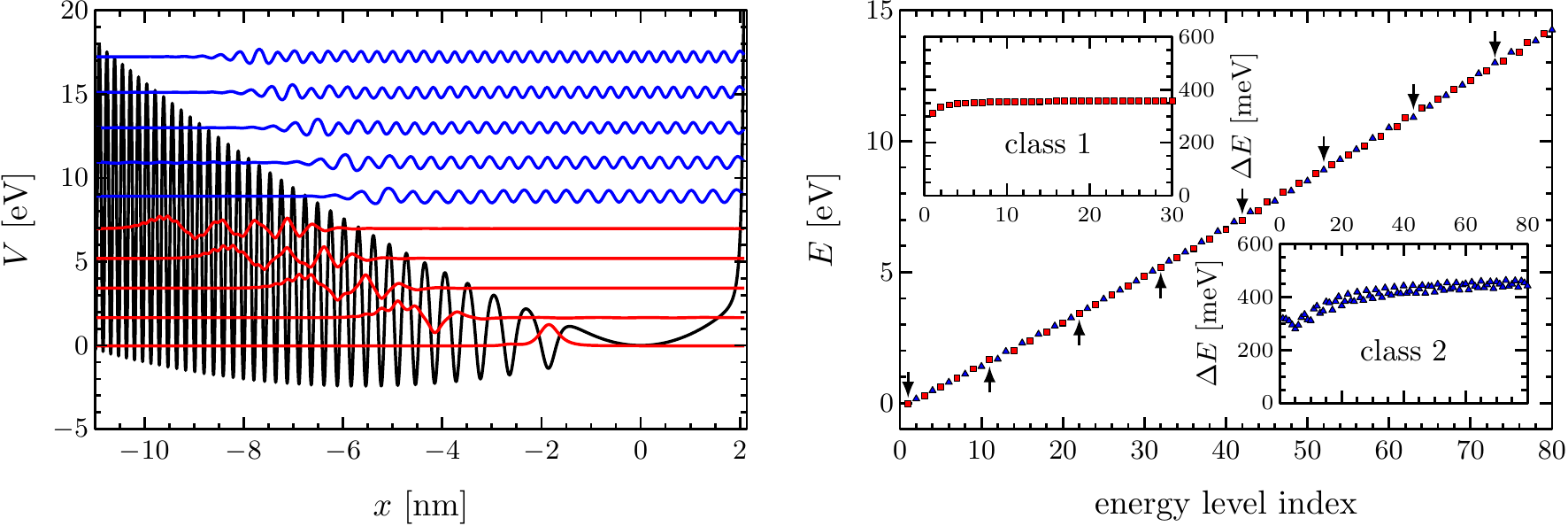}
	\caption{Numerical solution of \eqn{L3:U:DGL:2} with $A=-0.001$ and $W(1)=W'(1)=W''(1)=0$. Left: type-2 potential and selected states (quantum numbers 1, 11, 22, 32, 42, 52, 63, 73, 84, 95). The offset of the states are the corresponding eigenenergies. Two different classes of states are denoted by color (class 1: red, class 2: blue). Right: energy levels. Arrows mark the states shown in (a). The insets show the energy level spacings.}
	\label{fig:L3:num_2}
\end{figure}

\begin{figure}[t]
	\includegraphics[width=\TwoColumnWidth]{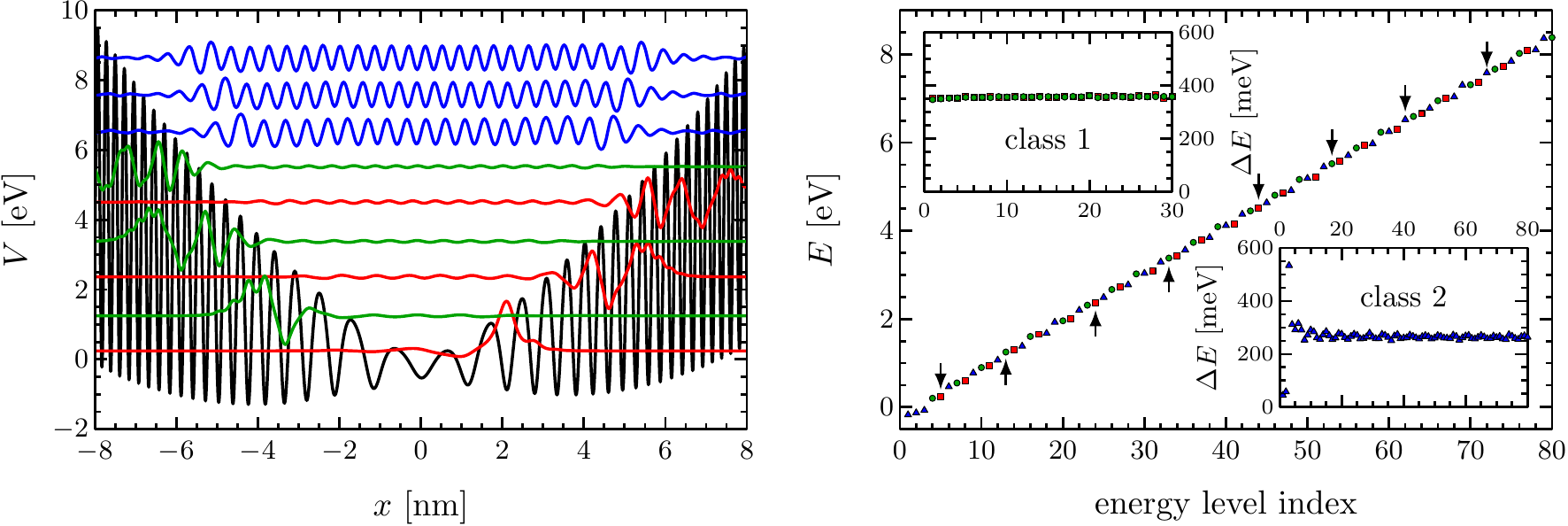}
	\caption{Numerical solution of \eqn{L3:U:DGL:3} with $A=0$, $B=-1$, and $W(1)=W'(1)=0$. Left: type-3 potential and selected states (quantum numbers 5, 13, 24, 33, 44, 53, 62, 72, 82). The offset of the states are the corresponding eigenenergies. Two different classes of states are denoted by color (class 1: red and green, class 2: blue). Right: energy levels. Arrows mark the states shown in (a). The insets show the energy level spacings (red data points are behind green data points).}
	\label{fig:L3:num_3}
\end{figure}

Three different potentials and corresponding solutions of the Schr\"odinger equation are depicted in Fig.~\ref{fig:L3:num_1}, Fig.~\ref{fig:L3:num_2}, and Fig.~\ref{fig:L3:num_3}.
They represent three different types of solutions, named type-1, type-2, and type-3 potentials in the following.

\fig{L3:num_1} shows a potential with two singularities and a minimum in between.
The singularities behave like $1/x^2$, similar to the $U_{2n+1}$ potential.
The wave functions are also of similar shape as the ones of the former potentials.
The energy levels are roughly equidistant in the depicted range with $\Delta E\approx 1040\,\text{meV}$.
The higher states increase quadratically in energy instead of linearly.
This could be due to the finite discretization and the resulting bad energetic resolution or due to a badly converged potential near the singularities.
However, the $1/x^2$ dependence near the singularities acts like a square box, resulting in quadratically increasing energy levels.

The second type of potentials, see example in \fig{L3:num_2}, also has a $1/x^2$ singularity for positive values of $x$, similar to $U_{2n+1}$.
Yet for $x<0$, the potential oscillates and the strength of these oscillations increases with rising absolute value of $x$.
The states of this potential can be separated into two classes:
(1) States localized in the oscillation minima (red states in \fig{L3:num_2}).
(2) Delocalized states oscillating between the smooth wall and the potential oscillation peaks of the corresponding energy range (blue states in \fig{L3:num_2}).
The depicted example exhibits nearly linearly increasing energy levels.
Additional small irregular variations are due to the 2 classes of states.
The insets in \fig{L3:num_2} show the energy level spacings of these distinct classes.
The class-1 states (upper inset) show linearly increasing energy levels with $\Delta E\approx 358\,\text{meV}$.
The class-2 states (lower inset) vary between $\Delta E=300\,\text{meV}$ and $\Delta E=450\,\text{meV}$.
The superposition of these two classes of states with different energy level spacings leads to the variations in the combined picture and although the class-2 states are not truly equidistant, the deviations from the overall linearity of all states are small.
The overall average energy level spacing is $\Delta E\approx 180\,\text{meV}$.

\Fig{L3:num_3} depicts the third type of potentials with no singularities, but oscillations in both directions and also displaying two different
classes of states.
Class 1 are states mostly localized within the potential oscillations at the left or right (red and green in \fig{L3:num_3}).
There are also states similar to the depicted ones, but located at the other side.
Class 2 are delocalized states oscillating between the left and right  potential oscillation peaks of corresponding energy (blue in \fig{L3:num_3}).
Energy levels increase almost linearly; the small periodic variations are due to the two classes of states.
The insets in \fig{L3:num_3} show the energy level spacings of these distinct classes.
It can be clearly seen that for each class linearly increasing energy levels are present.
Class 1 (upper inset in \fig{L3:num_3}) has $\Delta E=358\,\text{meV}$ for states located at the left (green) as well as for states located at the right (red).
The states are just shifted by a small energy due to the slightly asymmetry of the potential.
Class 2 (lower inset) has $\Delta E\approx 267\,\text{meV}$.
The class-2 energy level spacings vary slightly (about 10\%).
But as this is a differential value, deviations from linearity are sufficiently small and in good agreement with experimental results.
The superposition of these two classes of states with different energy level spacings leads to the irregular variations in the combined picture.
In total, the energy levels show a linear dependence with an overall average energy level spacing of $\Delta E\approx 108\,\text{meV}$.
The additional irregular variation on the large scale are small.
With regard to experimental results with some error bars these irregularities may not be visible.

The presented 1D potentials allow more interpretations concerning the comparison with real structures than a pure 1D harmonic oscillator model.
They can be related to different possible cases occurring in reality.
For example, a thin, free-standing slab, surrounded by vacuum, has states which are trapped within the slab in the direction perpendicular to the slab and exponentially decaying into vacuum.
This could be modeled by the type-1 potential of the numerical calculations, which diverges at the slab-vacuum interface.
A thin layer on a substrate could be modeled by the type-2 potential.
This diverges at the layer-vacuum interface and oscillates into the bulk-substrate, representing the periodic structure.
A thin sandwich-layer between two other substrates could be modeled by the type-3 potential, which oscillates into both directions to describe the penetration into the periodic bulk material.
Another interpretation could be to use the type-2 or type-3 potential to describe the oscillating atomic structure within the layer itself.
The experimental results from literature~\cite{PhysRevB.97.045403, PhysRevB.75.035422, PhysRevLett.115.106803} have the structure silicon--bismuth--vacuum and fall into these categories.
The $\Delta E=358\,\text{meV}$ achieved in \fig{L3:num_2} is in the range of possible experimental $\Delta E$ of \fig{HO:experiment}.
Using the result of the fit in \eqn{fit}, this corresponds to a thickness of $9.3$ bismuth bilayers, i.e. $3.7\,\text{nm}$.
Within the potential in \fig{L3:num_2} this approximately matches the width of the non-oscillating part at the right near the singularity.
The oscillating part represents the transition into the bulk-like part with wave lengths between $0.2\,\text{nm}$ (for $-10\,\text{nm}<x<-8\,\text{nm}$) and $0.6\,\text{nm}$ (for $-4\,\text{nm}<x<-2\,\text{nm}$).
This nicely matches the periodicity of possible atomic structures (the bismuth bilayer thickness is approximately $0.4\,\text{nm}$, the silicon lattice constant is $0.54\,\text{nm}$).

In this section spectra have been studied which are equidistant in a limited energy range or contain a subset of equidistant states.
In the next section we show what can be learned from perturbation theory.
The difference to the former shift-operator approach is that, in the following, searching for equidistant eigenstates means an infinite number of exactly equidistant states.
The formerly shown results are less strict and do not belong to this for the following reasons:
Potentials $U_{2m+1}$ and the numerical type-1 potential have non-polynomial shapes and quadratic non-integrable singularities.
Potentials $U_{2m}$ consider parts of the spectrum being non-equidistant (i.e. the ground state).
The numerical type-2 and type-3 potentials lead to states which split into two classes with separate energy spacing each.

\section{Perturbed 1D harmonic oscillator}\label{sec:Perturbation}

In this section, we use perturbation theory to show that an \textbf{exactly} equidistant energy spectrum \textbf{over the whole energy range} cannot be achieved by a polynomial potential other than the harmonic (quadratic) potential.
To this end, the 1D harmonic oscillator is perturbed with an arbitrary polynomial perturbation potential ${u(\xi)=\sum_{j=0}^Nu_j\xi^j}$,
\begin{align}
	\left[ -\dfrac{\partial^2}{\partial\xi^2} + \xi^2 + \lambda u(\xi) \right] \psi_n(\xi) &= \epsilon_n^{(0)}\psi_n(\xi).
\end{align}
As the equations are linear, we consider each monomial $\xi^j$ individually.
The perturbation potential ${u(\xi)=\xi^j}$ in the basis of the harmonic oscillator eigenstates gives the perturbation coefficients
\begin{align}
	u_{mn}^{(j)} &= \frac{1}{\sqrt{\pi 2^{m+n}m!n!}} \int\limits_{-\infty}^\infty \xi^j\text{e}^{-\xi^2}H_m(\xi)H_n(\xi) \,\text{d}\xi \quad,
\end{align}
where $H_n(\xi)$ are the Hermite polynomials.

\begin{figure}[t]
	\centering
	\includegraphics[width=\OneColumnWidth]{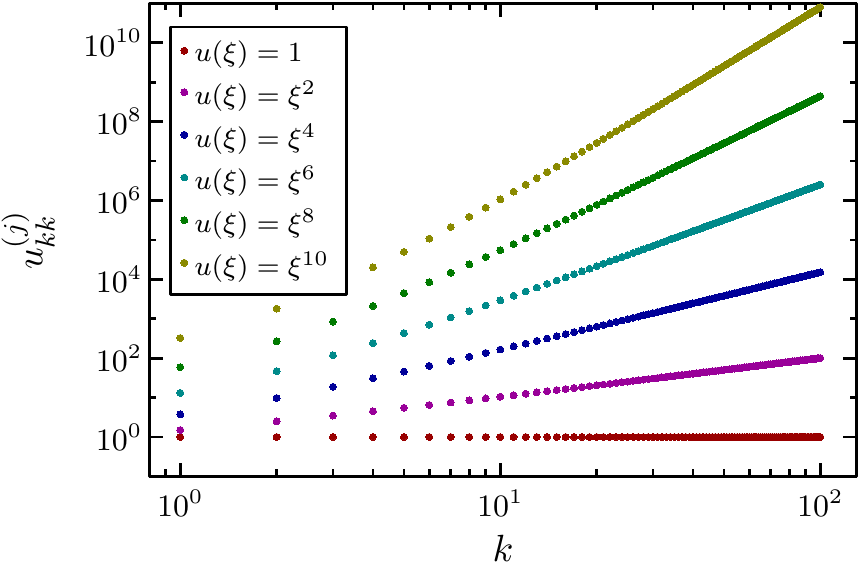}
	\caption{$u_{kk}^{(j)}$ for different perturbation potentials $u(\xi)=\xi^j$.}
	\label{fig:u_kk_j}
\end{figure}

The diagonal elements $u_{kk}^{(j)}$ equal the first-order energy correction $\epsilon_k^{(1)}$.
They are equal to zero for odd values of $j$.
For even numbers $j$, $u_{kk}^{(j)}$ is a polynomial in $k$ of order $j/2$ as shown in \fig{u_kk_j} for different monomial perturbations ${u(\xi)=\xi^j}$.
For the smallest values of $j$ one gets
\begin{align}
	u_{kk}^{(0)} = 1 \quad,\quad
	u_{kk}^{(2)} = \frac{1}{2} + k \quad,\quad
	u_{kk}^{(4)} = \frac{3}{4} + \frac{3}{2}k + \frac{3}{2}k^2 \quad.
\end{align}
Consequently, if the harmonic oscillator is perturbed with a quadratic potential, $\epsilon_k^{(1)}$ is linear in $k$, reproducing the equidistant energy levels of the harmonic oscillator.
If the harmonic oscillator is perturbed with a non-quadratic polynomial potential, $\epsilon_k^{(1)}$ is a superposition of different powers of $k$, yielding a different spectrum.
Thus, only the quadratic polynomial gives equidistant first-order energy corrections.

\begin{figure}[t]
	\includegraphics[width=\TwoColumnWidth]{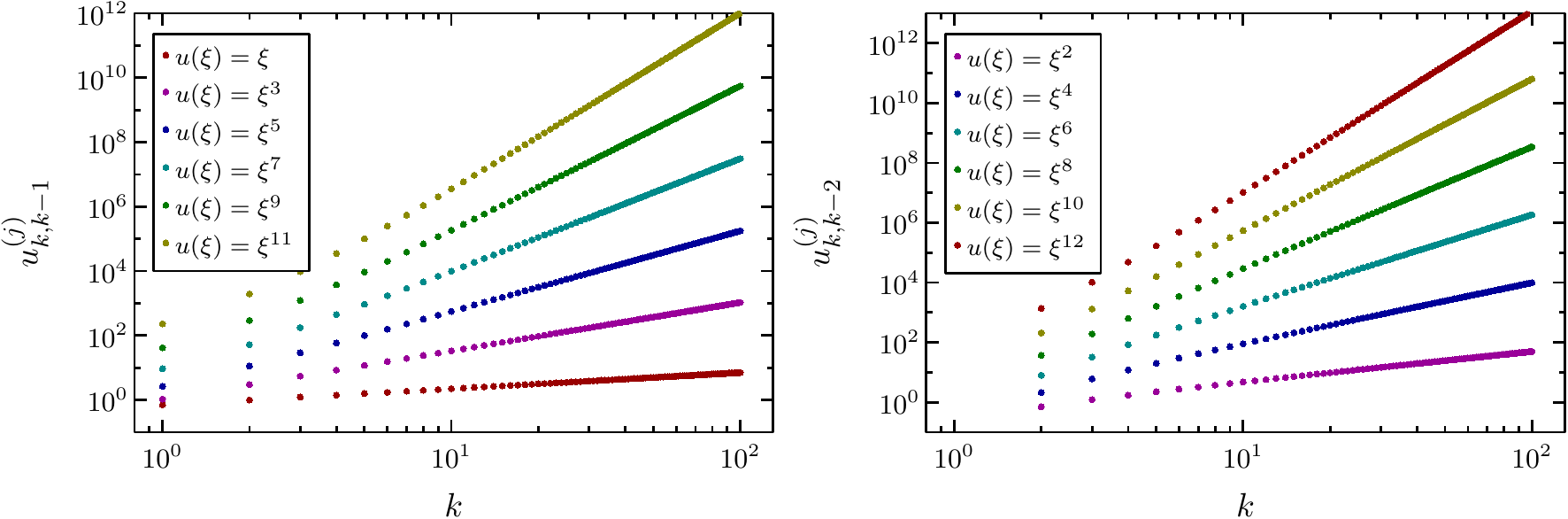}
	\caption{$u_{k,k-l}^{(j)}$ for different $l$ and for different perturbation potentials $u(\xi)=\xi^j$.}
	\label{fig:u_mn_j}
\end{figure}

The non-diagonal elements $u_{mn}^{(j)}$ are needed to calculate the higher order energy corrections $\epsilon_k^{(\geq 2)}$.
For $p$th order a $(p-1)$-fold sum over a $p$-fold product of $u_{mn}^{(j)}$ with different indices has to be calculated.
If $m+n+j$ is odd or if $|m-n|>j$, then ${u_{mn}^{(j)}=0}$.
This leads to finite summations over no more than $j+1$ elements $u_{k,k+l}^{(j)}$ with $-j\leq l\leq j$.
The $u_{k,k+l}^{(j)}$ are shown in \fig{u_mn_j} for different $l$ and for different polynomials ${u(\xi)=\xi^j}$.
The non-zero elements for a specific $l$ can be written as
\begin{align}
	u_{k,k-l}^{(j)} = \sqrt{\frac{k!}{2^l(k-l)!}}p_{kl}^{(j)}
\end{align}
with a polynomial $p_{kl}^{(j)}$ in $k$ of order $(j-l)/2$.
Thus, $(u_{k,k-l}^{(j)})^2$ is a polynomial in $k$ of order $j$ for all $l$.
The previous diagonal case $l=0$ is included.
In particular, for the smallest values of $j$ one gets
\begin{flalign}
	p_{k,1}^{(1)} &= 1 \quad,
	& p_{k,1}^{(3)} &= \frac{3}{2}k \quad,
	& p_{k,1}^{(5)} &= \frac{5}{4}(1+2k^2) \quad,\\
	p_{k,2}^{(2)} &= 1 \quad,
	& p_{k,2}^{(4)} &= -1+2k \quad,
	& p_{k,2}^{(6)} &= \frac{15}{4}(1-k+k^2) \quad.&&\hspace{15em}
\end{flalign}

With this, the second order energy correction
\begin{align}
	\epsilon_k^{(2)} &= -\sum_{n\neq k}\frac{u_{nk}u_{kn}}{\epsilon_n^{(0)}-\epsilon_k^{(0)}} = -\frac{1}{2} \sum_{\substack{1\leq l\leq j\\l-j~\text{even}}}\frac{(u_{k,k+l}^{(j)})^2-(u_{k,k-l}^{(j)})^2}{l} = \mathcal{O}(k^{j-1})
\end{align}
is a polynomial in $k$ of order $j-1$, as the highest order in $k$ cancels out with the difference $(u_{k,k+l}^{(j)})^2-(u_{k,k-l}^{(j)})^2$.
Consequently, the second order energy corrections scale linearly with $k$ only for the quadratic potential ($j=2$).
Treating the higher-order perturbations similarly, in the $k^{3j/2}$-scaling two hierarchic differences lead to a cancellation of the two highest orders of $k$.
With this, $\epsilon_k^{(3)}$ is a polynomial in $k$ of order $3j/2-2$ and also linear scaling only for the quadratic potential.

Another argumentation can be done using the $p$th order perturbation corrections, written as a recursion:
\begin{align}
	\epsilon_k^{(p)} &= \sum_nu_{kn}c_{kn}^{(p-1)} - \sum_{q=2}^{p-1}\epsilon_k^{(q)}c_{kk}^{(p-q)} \\
	c_{km}^{(p)} &= \frac{1}{\epsilon_k^{(0)}-\epsilon_m^{(0)}}\left( \sum_nu_{mn}c_{kn}^{(p-1)} - u_{kk}c_{km}^{(p-1)} - \sum_{q=2}^{p-1}\epsilon_k^{(q)}c_{km}^{(p-q)} \right) \quad.
\end{align}
The general argumentation is as follows:
For the harmonic oscillator, $c_{km}^{(0)}$ is independent of $k$ and $\epsilon_k^{(0)}$ is linear in $k$.
If $c_{km}^{(p-1)}$ is independent of $k$ and $\epsilon_k^{(p-1)}$ is linear in $k$, which is fulfilled for $p=1$, then $c_{km}^{(p)}$ is also independent of $k$ and $\epsilon_k^{(p)}$ is also linear in $k$.
For the case $j\neq 2$ yields: if $c_{km}^{(p-1)}$ scales like $k^x$ and $\epsilon_k^{(p-1)}$ scales like $k^y$, then $c_{km}^{(p)}$ scales like $k^{x+j/2-1}+k^{x+y-1}$ and $\epsilon_k^{(p)}$ scales like $k^{x+j/2}+k^{x+y}$.
Consequently, a linear dependence of $\epsilon_k$ on $k$ can only be achieved if $y=1$, thus $x=0$, giving $j=2$, which is the harmonic oscillator.
In total, for a general perturbation potential ${u(\xi)=\sum_{j=0}^Nu_j\xi^j}$, the perturbed energy levels are a sum of linear functions of $k$ for the quadratic case and a sum of functions of $k$ with different exponents for the non-quadratic cases, which does not sum up to a linear dependence.

Though it may seem so, \Sec{Shift_operator} and \Sec{Perturbation} do not contradict each other. 
The previous argumentation holds only for a polynomial potential which produces an equidistant spectrum for all eigenstates and with no tolerance in the level spacing.
It does not apply to power series and spectra which are only equidistant in a specific energy range or approximately equidistant spectra, as required to model experimental spectra.

The derivations of this section and some more results can be found in the Supplementary Material.

\section{Summary and conclusions}

In this publication, we discuss the implications that may be drawn from the experimental observation of an equidistant energy level spacing. 
Though a harmonic potential comes to mind immediately, this shape of potential is sufficient, but not necessary for the equidistant spacing, especially considering that experimental
results cover a finite energy range with a finite measurement uncertainty. The question is thus more involved than is visible at first sight.
In this publication, we discuss different aspects of the problem.
 
In the first part we compare a 1D harmonic oscillator model with experimental results, looking at the analytical harmonic potential as well as on
a truncated version. 
The energy levels measured in thin Bi films of different thicknesses and the resulting thickness-dependence fit the theoretical model with comparable energy values and dimensions.
Thus, the 1D harmonic oscillator is a valid model to describe the quantum well states in thin films as shown for Bi films.

However, this still leaves the question whether it is the only possible potential to explain the measurements.
Therefore, in the second part of the paper, we discuss anharmonic potentials of non-polynomial shape and apply the shift-operator 
approach to explicitly calculate potentials with an equidistant spectrum.
We present a number of analytic and numeric examples representing different types of potentials including asymmetric minima, singularities, and oscillations.
These potentials seem well-suited 1D models to describe the potential of thin slabs, thin films on substrates or thin sandwich films between two substrates.
They also fit the expectation that first principle calculations of real materials should result in oscillating potentials which are in average quadratic to fit the harmonic oscillator model.

In the last part, we apply $n$th-order perturbation theory with a polynomial perturbation potential to the quantum-mechanical harmonic oscillator problem.
We calculate and discuss the perturbation integrals $u_{k,k-l}$ and the resulting $n$th order energy corrections as a function of $k$ for single monomials $\xi^j$,
showing that $u_{k,k-l}^2$ is a polynomial in $k$ of order $j$ and arguing that every non-quadratic perturbation monomial leads to a non-quadratic $k$-dependence of the $n$th-order energy corrections, leaving only the harmonic term to produce an equidistant spectrum.

Thus we conclude that all possible anharmonic potentials producing exactly equidistant spectra must have a non-polynomial form.
Anharmonic non-polynomial potentials exist, which could describe real structures and produce spectra which mirror the experimental results.

\section*{Acknowledgment}

We thank C. Tegenkamp and H.-R. Berger for helpful discussions.


\begin{thebibliography}{\bstindent}

\bibitem{SciRep.5.17424}
{\bstauthor C.~Ke}, {\bstauthor W.~Zhu}, {\bstauthor Z.~Zhang}, {\bstauthor
  E.~Soon~Tok}, {\bstauthor B.~Ling}, \bbland~{\bstauthor J.~Pan}: {\bsttitle
  Thickness-Induced Metal-Insulator Transition in Sb-doped SnO$_2$ Ultrathin
  Films: The Role of Quantum Confinement},
  \href{http://dx.doi.org/10.1038/srep17424}{{\bstjournal Scientific Reports}
  {\bstvolume 5} ({\bstyear 2015}), 17424}.

\bibitem{PhysStatSolB.33.425}
{\bstauthor K.~Berchtold}, \bbland~{\bstauthor D.~Huber}: {\bsttitle Transport
  Properties of Indium Antimonide Thin Films},
  \href{http://dx.doi.org/10.1002/pssb.19690330142}{{\bstjournal physica status
  solidi (b)} {\bstvolume 33} ({\bstyear 1969}), 425--429}.

\bibitem{PhysRevB.66.233408}
{\bstauthor C.~M. Wei}, \bbland~{\bstauthor M.~Y. Chou}: {\bsttitle Theory of
  quantum size effects in thin Pb(111) films},
  \href{http://dx.doi.org/10.1103/PhysRevB.66.233408}{{\bstjournal Physical
  Review B} {\bstvolume 66} ({\bstyear 2002}), 233408}.

\bibitem{PhysRevB.97.045403}
{\bstauthor P.~Kr\"oger}, {\bstauthor D.~Abdelbarey}, {\bstauthor M.~Siemens},
  {\bstauthor D.~L\"ukermann}, {\bstauthor S.~Sologub}, {\bstauthor
  H.~Pfn\"ur}, \bbland~{\bstauthor C.~Tegenkamp}: {\bsttitle Controlling
  conductivity by quantum well states in ultrathin Bi(111) films},
  \href{http://dx.doi.org/10.1103/PhysRevB.97.045403}{{\bstjournal Physical
  Review B} {\bstvolume 97} ({\bstyear 2018}), 045403}.

\bibitem{PhysRevB.75.035422}
{\bstauthor T.~Hirahara}, {\bstauthor T.~Nagao}, {\bstauthor I.~Matsuda},
  {\bstauthor G.~Bihlmayer}, {\bstauthor E.~V. Chulkov}, {\bstauthor Y.~M.
  Koroteev}, \bbland~{\bstauthor S.~Hasegawa}: {\bsttitle Quantum well states
  in ultrathin Bi films: Angle-resolved photoemission spectroscopy and
  first-principles calculations study},
  \href{http://dx.doi.org/10.1103/PhysRevB.75.035422}{{\bstjournal Physical
  Review B} {\bstvolume 75} ({\bstyear 2007}), 035422}.

\bibitem{PhysRevLett.115.106803}
{\bstauthor T.~Hirahara}, {\bstauthor T.~Shirai}, {\bstauthor T.~Hajiri},
  {\bstauthor M.~Matsunami}, {\bstauthor K.~Tanaka}, {\bstauthor S.~Kimura},
  {\bstauthor S.~Hasegawa}, \bbland~{\bstauthor K.~Kobayashi}: {\bsttitle Role
  of Quantum and Surface-State Effects in the Bulk Fermi-Level Position of
  Ultrathin Bi Films},
  \href{http://dx.doi.org/10.1103/PhysRevLett.115.106803}{{\bstjournal Physical
  Review Letters} {\bstvolume 115} ({\bstyear 2015}), 106803}.

\bibitem{PhysRevLett.93.046403}
{\bstauthor Y.~M. Koroteev}, {\bstauthor G.~Bihlmayer}, {\bstauthor J.~E.
  Gayone}, {\bstauthor E.~V. Chulkov}, {\bstauthor S.~Bl\"ugel}, {\bstauthor
  P.~M. Echenique}, \bbland~{\bstauthor P.~Hofmann}: {\bsttitle Strong
  Spin-Orbit Splitting on Bi Surfaces},
  \href{http://dx.doi.org/10.1103/PhysRevLett.93.046403}{{\bstjournal Physical
  Review Letters} {\bstvolume 93} ({\bstyear 2004}), 046403}.

\bibitem{PhysRevLett.117.236402}
{\bstauthor S.~Ito}, {\bstauthor B.~Feng}, {\bstauthor M.~Arita}, {\bstauthor
  A.~Takayama}, {\bstauthor R.-Y. Liu}, {\bstauthor T.~Someya}, {\bstauthor
  W.-C. Chen}, {\bstauthor T.~Iimori}, {\bstauthor H.~Namatame}, {\bstauthor
  M.~Taniguchi}, {\bstauthor C.-M. Cheng}, {\bstauthor S.-J. Tang}, {\bstauthor
  F.~Komori}, {\bstauthor K.~Kobayashi}, {\bstauthor T.-C. Chiang},
  \bbland~{\bstauthor I.~Matsuda}: {\bsttitle Proving Nontrivial Topology of
  Pure Bismuth by Quantum Confinement},
  \href{http://dx.doi.org/10.1103/PhysRevLett.117.236402}{{\bstjournal Physical
  Review Letters} {\bstvolume 117} ({\bstyear 2016}), 236402}.

\bibitem{JElectronSpec.201.98}
{\bstauthor T.~Hirahara}: {\bsttitle The Rashba and quantum size effects in
  ultrathin Bi films},
  \href{http://dx.doi.org/https://doi.org/10.1016/j.elspec.2014.08.004}{{\bstjournal
  Journal of Electron Spectroscopy and Related Phenomena} {\bstvolume 201}
  ({\bstyear 2015}), 98--104}.

\bibitem{NanoLett.12.1776}
{\bstauthor A.~Takayama}, {\bstauthor T.~Sato}, {\bstauthor S.~Souma},
  {\bstauthor T.~Oguchi}, \bbland~{\bstauthor T.~Takahashi}: {\bsttitle Tunable
  Spin Polarization in Bismuth Ultrathin Film on Si(111)},
  \href{http://dx.doi.org/10.1021/nl2035018}{{\bstjournal Nano Letters}
  {\bstvolume 12} ({\bstyear 2012}), 1776--1779}.

\bibitem{CommMathPhys.82.471}
{\bstauthor H.~P. McKean}, \bbland~{\bstauthor E.~Trubowitz}: {\bsttitle The
  spectral class of the quantum-mechanical harmonic oscillator},
  \href{https://projecteuclid.org/euclid.cmp/1103920654}{{\bstjournal
  Communications in Mathematical Physics} {\bstvolume 82} ({\bstyear 1981}),
  471--495}.

\bibitem{JMathPhys.25.3387}
{\bstauthor B.~Mielnik}: {\bsttitle Factorization method and new potentials
  with the oscillator spectrum},
  \href{http://dx.doi.org/10.1063/1.526108}{{\bstjournal Journal of
  Mathematical Physics} {\bstvolume 25} ({\bstyear 1984}), 3387--3389}.

\bibitem{TheorChemAcc.110.403}
{\bstauthor J.~Morales}, {\bstauthor J.~J. Pe{\~{n}}a}, \bbland~{\bstauthor
  A.~Rubio-Ponce}: {\bsttitle New isospectral generalized potentials},
  \href{http://dx.doi.org/10.1007/s00214-003-0494-7}{{\bstjournal Theoretical
  Chemistry Accounts} {\bstvolume 110} ({\bstyear 2003}), 403--409}.

\bibitem{Chaos.4.47}
{\bstauthor S.~Y. Dubov}, {\bstauthor V.~M. Eleonskii}, \bbland~{\bstauthor
  N.~E. Kulagin}: {\bsttitle Equidistant spectra of anharmonic oscillators},
  \href{http://dx.doi.org/10.1063/1.166056}{{\bstjournal Chaos: An
  Interdisciplinary Journal of Nonlinear Science} {\bstvolume 4} ({\bstyear
  1994}), 47--53}.

\bibitem{SovPhysJETP.75.446}
{\bstauthor S.~Y. Dubov}, {\bstauthor V.~M. Eleonskii}, \bbland~{\bstauthor
  N.~E. Kulagin}: {\bsttitle Equidistant spectra of anharmonic oscillators},
  \href{http://www.jetp.ac.ru/cgi-bin/e/index/e/75/3/p446?a=list}{{\bstjournal
  Journal of Experimental and Theoretical Physics} {\bstvolume 75} ({\bstyear
  1992}), 446--451}.

\bibitem{PhysLettA.70.177}
{\bstauthor Y.~Weissman}, \bbland~{\bstauthor J.~Jortner}: {\bsttitle The
  isotonic oscillator},
  \href{http://dx.doi.org/https://doi.org/10.1016/0375-9601(79)90197-X}{{\bstjournal
  Physics Letters A} {\bstvolume 70} ({\bstyear 1979}), 177--179}.

\bibitem{IntJQuantumChem.110.1317}
{\bstauthor A.~R. Matamala}, {\bstauthor C.~A. Salas}, \bbland~{\bstauthor
  J.~F. Cari�ena}: {\bsttitle Degeneracy in one-dimensional quantum mechanics:
  A case study}, \href{http://dx.doi.org/10.1002/qua.22225}{{\bstjournal
  International Journal of Quantum Chemistry} {\bstvolume 110} ({\bstyear
  2010}), 1317--1321}.

\bibitem{JPhysAMathGen.20.4331}
{\bstauthor D.~Zhu}: {\bsttitle A new potential with the spectrum of an
  isotonic oscillator},
  \href{http://dx.doi.org/10.1088/0305-4470/20/13/034}{{\bstjournal Journal of
  Physics A: Mathematical and General} {\bstvolume 20} ({\bstyear 1987}),
  4331--4336}.

\bibitem{JPhysAMathTheo.41.085301}
{\bstauthor J.~F. Cari{\~{n}}ena}, {\bstauthor A.~M. Perelomov}, {\bstauthor
  M.~F. Ra{\~{n}}ada}, \bbland~{\bstauthor M.~Santander}: {\bsttitle A quantum
  exactly solvable nonlinear oscillator related to the isotonic oscillator},
  \href{http://dx.doi.org/10.1088/1751-8113/41/8/085301}{{\bstjournal Journal
  of Physics A: Mathematical and Theoretical} {\bstvolume 41} ({\bstyear
  2008}), 085301}.

\bibitem{OsakaJMath.36.949}
{\bstauthor M.~Ohmiya}: {\bsttitle Spectrum of Darboux transformation of
  differential operator}, \href{http://dx.doi.org/10.18910/6851}{{\bstjournal
  Osaka Journal of Mathematics} {\bstvolume 36} ({\bstyear 1999}), 949--980}.

\bibitem{Mathematica.12.1}
{\bsttitle {Mathematica, {V}ersion 12.1, Wolfram Research, Inc.}},
  \href{https://www.wolfram.com/mathematica}{https://www.wolfram.com/mathematica}.

\end{thebibliography}
\end{document}


\maketitle

\section{Shift-operator approach}

\subsection{Definition and basic operator equation}

\noindent The Schr\"odinger equation in dimensionless coordinates
\begin{align}
\begin{aligned}
	\xi &= \sqrt{\frac{m\omega}{\hbar}}x \quad,& \epsilon_n &= \frac{E_n}{\hbar\omega} \quad,& \psi(\xi) &= \sqrt[4]{\frac{\hbar}{m\omega}}\Psi(x)\quad,& U(\xi) &= \frac{V(x)}{\hbar\omega}
\end{aligned}
\end{align}
is
\begin{align}
	\left[ -\dfrac{1}{2}\dfrac{\partial^2}{\partial\xi^2} + U(\xi) \right] \psi_n(\xi) = \epsilon_n\psi_n(\xi) \quad .
\end{align}
The shift operator
\begin{align}
	\mathcal{L}\psi(\xi,\epsilon) = \psi(\xi,\epsilon+1) \quad,\quad \mathcal{L}^\dagger\psi(\xi,\epsilon) = \psi(\xi,\epsilon-1)
\end{align}
shifts each state in energy by $\hbar\omega$.
Applying the Hamiltonian operator gives
\begin{align}
\begin{aligned}
	\mathcal{H}\mathcal{L}\psi(\xi,\epsilon) &= (\epsilon+1)\psi(\xi,\epsilon+1) \quad,& \mathcal{H}\mathcal{L}^\dagger\psi(\xi,\epsilon) &= (\epsilon-1)\psi(\xi,\epsilon-1) \quad,\\
	\mathcal{L}\mathcal{H}\psi(\xi,\epsilon) &= \epsilon\psi(\xi,\epsilon+1) \quad,& \mathcal{L}^\dagger\mathcal{H}\psi(\xi,\epsilon) &= \epsilon\psi(\xi,\epsilon-1) \quad.
\end{aligned}
\end{align}
For an equidistant spectrum $\epsilon_{n+1}=\epsilon_n+1$ this gets
\begin{align}
\begin{aligned}
	\mathcal{H}\mathcal{L}\psi_n(\xi) &= \epsilon_{n+1}\psi_{n+1}(\xi) \quad,& \mathcal{H}\mathcal{L}^\dagger\psi_n(\xi) &= \epsilon_{n-1}\psi_{n-1}(\xi) \quad,\\
	\mathcal{L}\mathcal{H}\psi_n(\xi) &= \epsilon_n\psi_{n+1}(\xi) \quad,& \mathcal{L}^\dagger\mathcal{H}\psi_n(\xi) &= \epsilon_n\psi_{n-1}(\xi) \quad.
\end{aligned}
\end{align}
With this, a resulting operator equation can be set up using commutators:
\begin{align}
	[\mathcal{H},\mathcal{L}] = \mathcal{L} \quad,\quad [\mathcal{H},\mathcal{L}^\dagger] &= -\mathcal{L}^\dagger \quad.\label{eqn:shift_operator_equation}
\end{align}
It follows $[\mathcal{H},\mathcal{L}\mathcal{L}^\dagger] = [\mathcal{H},\mathcal{L}^\dagger\mathcal{L}] = 0$.
Consequently, $\mathcal{L}\mathcal{L}^\dagger$, $\mathcal{L}^\dagger\mathcal{L}$, and $[\mathcal{L},\mathcal{L}^\dagger]$ are functions of $\mathcal{H}$. Thus, $\mathcal{L}$ can be written as a polynomial of the dimensionless momentum operator $\mathcal{P}=-\text{i}\frac{\partial}{\partial\xi}$:
\begin{align}
	\mathcal{L} = \sum_{k=0}^K \alpha_k(\xi)(\text{i}\mathcal{P})^k \quad.\label{eqn:L:general}
\end{align}
$K$ is the order of the shift operator.
$\alpha_k(\xi)$ are free parameters.
Inserting \eqn{L:general} into \eqn{shift_operator_equation} and comparing the coefficients yields $K+2$ differential equations for $K+1$ unknown functions $\alpha_k(\xi)$ and the additional unknown function $U(\xi)$.

\subsection{First-order shift operator}

\begin{figure}
	\centering
	\includegraphics[width=\OneColumnWidth]{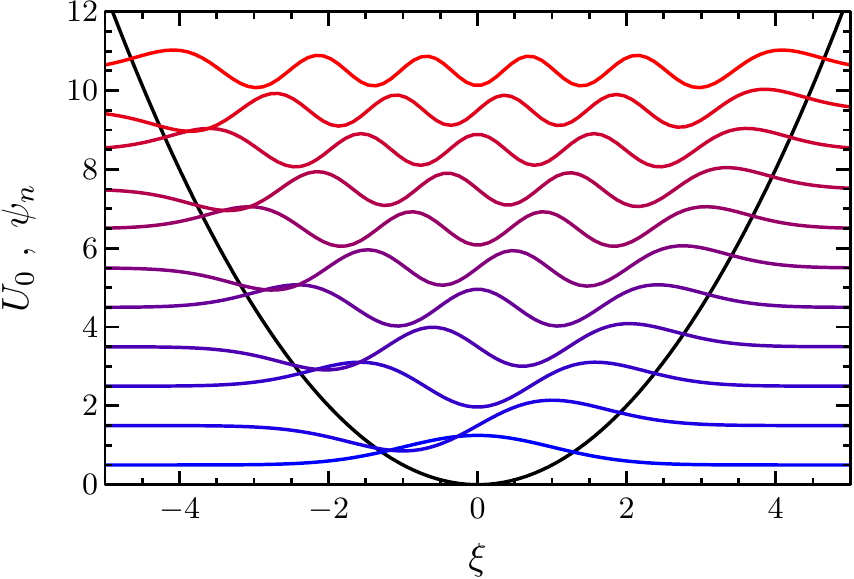}
	\caption{Potential and lowest states of the first-order shift operator.}
	\label{fig:L1}
\end{figure}

\noindent Inserting the first-order shift operator
\begin{align}
	\mathcal{L} = \alpha_0(\xi) + \alpha_1(\xi)\text{i}\mathcal{P}
\end{align}
into \eqn{shift_operator_equation} yields
\begin{alignat}{4}
\begin{aligned}
	0 &= \alpha_0 + \frac{1}{2}\alpha_0'' + U'\alpha_1 \quad,\\
	0 &= \alpha_0' + \alpha_1 + \frac{1}{2}\alpha_1'' \quad,\\
	0 &= \alpha_1' \quad.
\end{aligned}
\end{alignat}
The solution for the $\alpha_k$ yields the condition
\begin{align}
	U' = \xi \quad.
\end{align}
Integrating this yields the harmonic oscillator potential
\begin{align}
	U_0 = \frac{1}{2}\xi^2 \quad.
\end{align}
The corresponding shift operator
\begin{align}
	\mathcal{L} = -\xi + \text{i}\mathcal{P} \quad,\quad \mathcal{L}^\dagger = \xi + \text{i}\mathcal{P}
\end{align}
is the creation resp.~the annihilation operator.
The corresponding solution of the Schr\"odinger equation is
\begin{align}
	\epsilon_n &= \frac{1}{2} + n \quad,\quad \psi_n(\xi) = \frac{1}{\sqrt{2^nn!\sqrt{\pi}}} H_n(\xi)\exp\left(-\frac{1}{2}\xi^2\right) \quad,
\end{align}
where $H_n(\xi)$ are the Hermite polynomials.
The potential and the states of the first-order shift operator are depicted in \fig{L1}.

\subsection{Second-order shift operator}

\begin{figure}
	\centering
	\includegraphics[width=\OneColumnWidth]{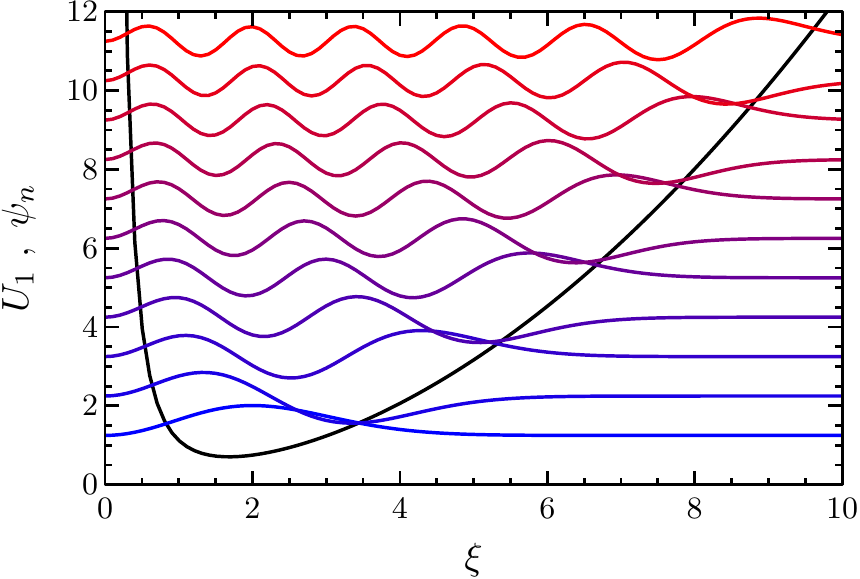}
	\caption{Potential and lowest states of the second-order shift operator.}
	\label{fig:L2}
\end{figure}

\noindent Inserting the second-order shift operator
\begin{align}
	\mathcal{L} = \alpha_0(\xi) + \alpha_1(\xi)\text{i}\mathcal{P} + \alpha_2(\xi)(\text{i}\mathcal{P})^2
\end{align}
into \eqn{shift_operator_equation} yields
\begin{alignat}{5}
\begin{aligned}
	0 &= \alpha_0 + \frac{1}{2}\alpha_0'' + U'\alpha_1 + U''\alpha_2 = 0 \quad,\\
	0 &= \alpha_0' + \alpha_1 + \frac{1}{2}\alpha_1'' + 2U'\alpha_2 \quad,\\
	0 &= \alpha_1' + \alpha_2 + \frac{1}{2}\alpha_2'' \quad,\\
	0 &= \alpha_2' \quad.
\end{aligned}
\end{alignat}
The solution for the $\alpha_k$ yields the condition
\begin{align}
	\xi U' + 2U = \frac{1}{2}\xi^2 \quad.
\end{align}
Multiplying with $\xi$ and summarizing using the product rule yields
\begin{align}
	(\xi^2U)' = \frac{1}{2}\xi^3 \quad.
\end{align}
Integrating and dividing by $\xi^2$ yields
\begin{align}
	U_1 = \frac{1}{8}\xi^2 + \frac{A}{\xi^2} \quad.
\end{align}
The shift operator is
\begin{align}
\begin{aligned}
	\mathcal{L} &= \frac{1}{4}\xi^2 - \frac{2A}{\xi^2} - \frac{1}{2} - \xi\text{i}\mathcal{P} - (\text{i}\mathcal{P})^2 \quad,\quad \mathcal{L}^\dagger = \frac{1}{4}\xi^2 - \frac{2A}{\xi^2} + \frac{1}{2} + \xi\text{i}\mathcal{P} - (\text{i}\mathcal{P})^2 \quad.
\end{aligned}
\end{align}
The solution of the Schr\"odinger equation can be achieved with the asymptotic behaviour
\begin{align}
	\xi &\rightarrow\infty: & \psi_\infty''-\frac{1}{2}\xi^2\psi_\infty &= 0 \quad, & \psi_\infty &\sim \exp\left(-\frac{1}{4}\xi^2\right) \quad,\hspace{20em}\\
	\xi &\rightarrow 0: & \psi_0''-\frac{4A}{\xi^2}\psi_0 &= 0 \quad, & \psi_0 &\sim \xi^{(1+\sqrt{1+8A})/2}
\end{align}
and the consequential approach
\begin{align}
	\psi_n(\xi) &= \xi^{(1+\sqrt{1+8A})/2}\sum_{k=0}^na_k\xi^k\exp\left(-\frac{1}{4}\xi^2\right) \quad.
\end{align}
It follows
\begin{align}
	\epsilon_n &= \frac{1}{2} + n + \frac{1}{4}\sqrt{1+8A} \quad,\\
	\psi_n(\xi) &= C_n J_n(\xi) \left(\frac{\xi^2}{2}\right)^{(1+\sqrt{1+8A})/4} \exp\left(-\frac{1}{4}\xi^2\right) \quad,\\
	J_n(\xi) &= \sum_{k=0}^n (-1)^k \begin{pmatrix}n\\k\end{pmatrix} \frac{\Gamma\left(\frac{\sqrt{1+8A}}{2}+1\right)}{\Gamma\left(\frac{\sqrt{1+8A}}{2}+1+k\right)} \left(\frac{\xi^2}{2}\right)^k \quad,\\
	C_n &= \sqrt{\frac{\sqrt{2}\Gamma\left(\frac{\sqrt{1+8A}}{2}+1+n\right)}{n!\Gamma^2\left(\frac{\sqrt{1+8A}}{2}+1\right)}} \quad.
\end{align}
The potential and the states of the second-order shift operator are depicted in \fig{L2}.

\subsection{Third-order shift operator}

\begin{figure}
	\centering
	\includegraphics[width=\TwoColumnWidth]{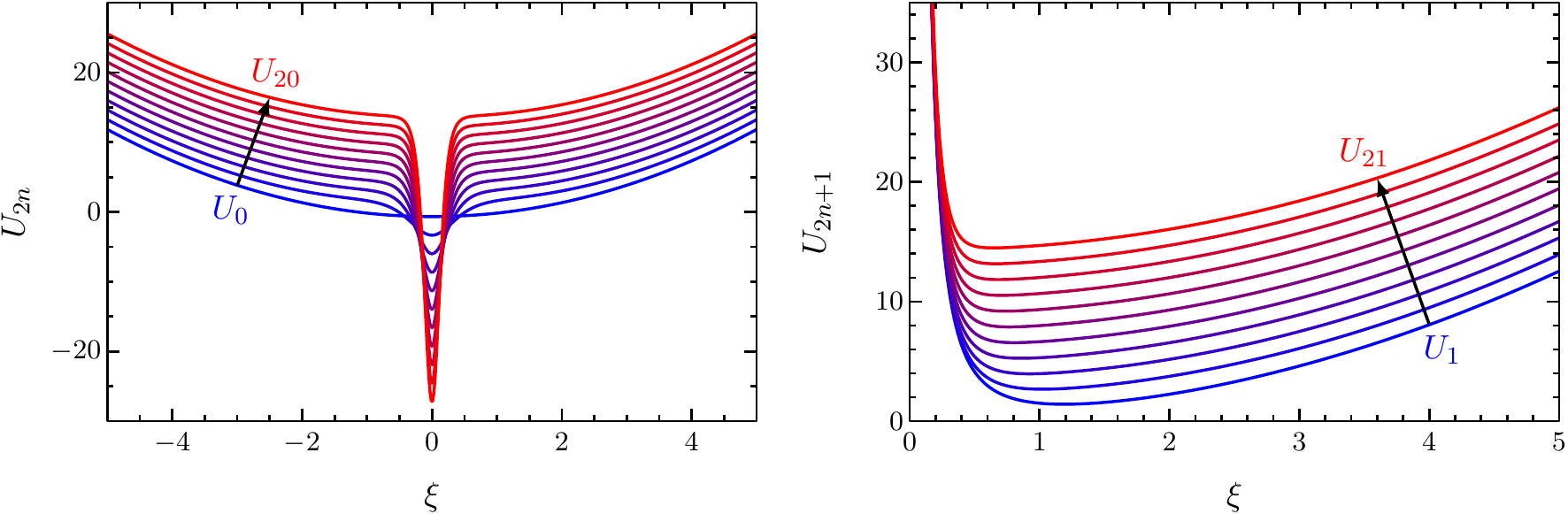}
	\caption{Potentials of the third-order shift operator obtained by a Darboux transformation.}
	\label{fig:L3:U}
\end{figure}

\begin{figure}[!b]
	\centering
	\begin{minipage}{\OneColumnWidth}
		\centering
		\includegraphics[width=\textwidth]{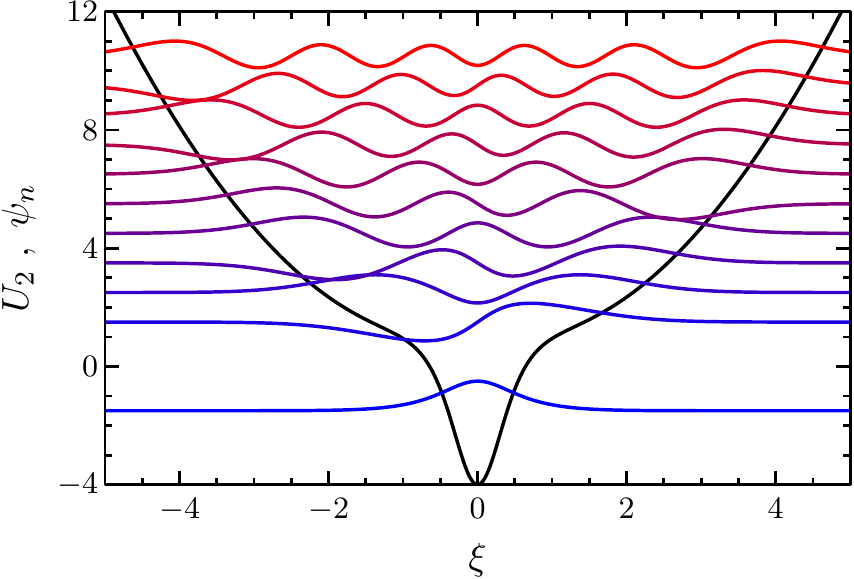}
		\caption{Potential $U_2$ and corresponding lowest states of the third-order shift operator.}
		\label{fig:L3:U2}
	\end{minipage}\hfill
	\begin{minipage}{\OneColumnWidth}
		\centering
		\includegraphics[width=\textwidth]{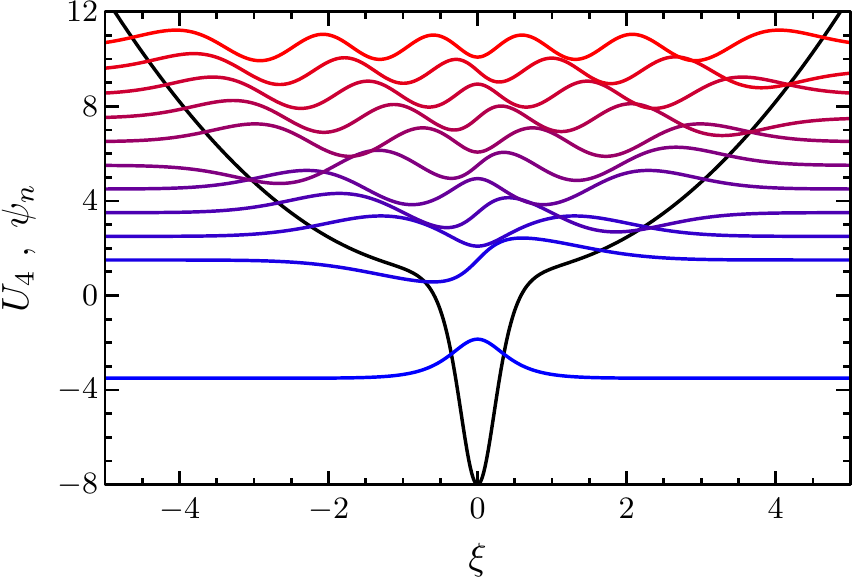}
		\caption{Potential $U_4$ and corresponding lowest states of the third-order shift operator.}
		\label{fig:L3:U4}
	\end{minipage}
\end{figure}

\noindent Inserting the thirt-order shift operator
\begin{align}
	\mathcal{L} = \alpha_0(\xi) + \alpha_1(\xi)\text{i}\mathcal{P} + \alpha_2(\xi)(\text{i}\mathcal{P})^2 + \alpha_3(\xi)(\text{i}\mathcal{P})^3
\end{align}
into \eqn{shift_operator_equation} yields
\begin{alignat}{6}
	0 &= \alpha_0 + \frac{1}{2}\alpha_0'' + U'\alpha_1 + U''\alpha_2 + U'''\alpha_3 \quad,\\
	0 &= \alpha_0' + \alpha_1 + \frac{1}{2}\alpha_1'' + 2U'\alpha_2 + 3U''\alpha_3 \quad,\\
	0 &= \alpha_1' + \alpha_2 + \frac{1}{2}\alpha_2'' + 3U'\alpha_3 \quad,\\
	0 &= \alpha_2' + \alpha_3 + \frac{1}{2}\alpha_3'' \quad,\\
	0 &= \alpha_3' \quad.
\end{alignat}
The solution for the $\alpha_k$ yields the shift operator
\begin{align}
	\mathcal{L} &= \left( \frac{3}{2}\left(U^2\right)' - \frac{1}{4}U''' - \frac{\xi^2+3}{2}U' + \frac{1}{2}\xi \right) + \left( \frac{1}{2}\xi^2 - 3U - 1 \right) \text{i}\mathcal{P} - \xi(\text{i}\mathcal{P})^2 + (\text{i}\mathcal{P})^3 \quad,\\
	\mathcal{L}^\dagger &= -\left( \frac{3}{2}\left(U^2\right)' - \frac{1}{4}U''' - \frac{\xi^2+3}{2}U' - \frac{1}{2}\xi \right) + \left( \frac{1}{2}\xi^2 - 3U + 1 \right) \text{i}\mathcal{P} + \xi(\text{i}\mathcal{P})^2 + (\text{i}\mathcal{P})^3
\end{align}
and the condition
\begin{align}
	\frac{3}{2}\left(W^2\right)'' - \frac{1}{4}W'''' + \xi^2W'' + 3\xi W' = 0
\end{align}
with $U = W + \frac{1}{2}\xi^2$.
Multiplying with $\xi$ and summarizing using the product rule yields the first integral
\begin{align}
	\frac{3}{2}\xi\left(W^2\right)' - \frac{3}{2}W^2 - \frac{1}{4}\xi W''' + \frac{1}{4}W'' + \xi^3W' = A \quad.\label{eqn:DGL:W}
\end{align}
Multiplying with $W'$, using the identities of \eqn{DGL:W}
\begin{gather}
	\frac{3}{2}\xi\left(W^2\right)' - \frac{1}{4}\xi W''' + \xi^3W' = A + \frac{3}{2}W^2 - \frac{1}{4}W'' \quad,\\
	 \xi^3W' = -\xi^2\left( \frac{A + \frac{3}{2}W^2 - \frac{1}{4}W''}{\xi} \right)' \quad,
\end{gather}
and summarizing using the chain rule yields another first integral
\begin{align}
	&-\frac{1}{2}\left( \frac{A + \frac{3}{2}W^2 - \frac{1}{4}W''}{\xi} \right)^2 - \frac{1}{2}W^3 + \frac{1}{8}\left(W'\right)^2 = AW + B \quad.
\end{align}
Using a Darboux transformation, a possible solution can be written in the form
\begin{align}
	W &= -\xi^2 + \frac{4}{3}\mu + \left( \frac{\chi_\mu'}{\chi_\mu} \right)^2
\end{align}
with
\begin{align}
	-\frac{1}{2}\chi_\mu'' + \frac{1}{2}\xi^2\chi_\mu = \mu\chi_\mu \quad.
\end{align}
With $\mu=-\left(m+\dfrac{1}{2}\right)$ two possible classes of solutions are
\begin{align}
	U_m &= -\frac{1}{2}\xi^2 - \frac{2}{3}(2m+1) + \left( \frac{P_m'(\xi)}{P_m(\xi)} + \xi \right)^2 \quad,\\
	P_{2m} &= \sum_{k=0}^m\frac{4^k}{(m-k)!(2k)!}\xi^{2k} \quad,\\
	P_{2m+1} &= \sum_{k=0}^m\frac{4^k}{(m-k)!(2k+1)!}\xi^{2k+1} \quad.
\end{align}
$U_{2n}$ are quadratic potentials with a dip at the minimum.
$U_0$ is the harmonic potential.
The corresponding energy levels are equidistant with $\Delta\epsilon=1$.
The ground state of potential $U_{2n}$ is additionally lowered by $2n$.
$U_{2n+1}$ are singular potentials at $\xi=0$.
$U_1$ is the one of the second-order shift operator.
The corresponding energy levels are equidistant with $\Delta\epsilon=2$ ($\Delta\epsilon=1$ can be achieved by rescaling $\xi=\xi'/2$).
The potentials of the third-order shift operator are depicted in \fig{L3:U}. The potentials $U_2$ and $U_4$ and corresponding lowest states are shown in \fig{L3:U2} and \fig{L3:U4}.

\section{Perturbed 1D harmonic oscillator}

\noindent The Schr\"odinger equation for the 1D harmonic oscillator is
\begin{align}
	\left[ -\frac{\hbar^2}{2m}\dfrac{\partial^2}{\partial x^2} + \frac{m\omega^2}{2}x^2 \right] \Psi_n(x) &= E_n^{(0)}\Psi_n(x) \quad.
\end{align}
Substituting with the dimensionless variables
\begin{align}
	x &= \sqrt{\frac{\hbar}{m\omega}}\xi \quad,\quad E_n^{(0)} = \hbar\omega\epsilon_n^{(0)} \quad,\quad \Psi(x) = \sqrt[4]{\frac{m\omega}{\hbar}}\psi(\xi)
\end{align}
yields the dimensionless formula
\begin{align}
	\left[ -\frac{1}{2}\dfrac{\partial^2}{\partial\xi^2} + \frac{1}{2}\xi^2 \right] \psi_n(\xi) &= \epsilon_n^{(0)}\psi_n(\xi) \quad.
\end{align}
Its solution is
\begin{align}
	\epsilon_n^{(0)} &= n+\frac{1}{2} \quad,\quad \psi_n(\xi) = \frac{1}{\sqrt{2^nn!\sqrt{\pi}}} \exp\left(-\frac{1}{2}\xi^2\right) H_n\left(\xi\right) \quad,
\end{align}
where $H_n$ are the Hermite polynomials. They can be expressed the following:
\begin{align}
	H_n(\xi) &= 2\xi H_{n-1}(\xi) - 2(n-1)H_{n-2}(\xi) \quad.\label{eqn:Hermite:recursive}
\end{align}

\noindent The Schr\"odinger equation of a system, which is perturbed with the perturbation potential $v(x) = \frac{1}{2}\hbar\omega u(\xi)$, respective the corresponding dimensionless perturbation potential $u(\xi)$, is
\begin{align}
	\left[ \mathcal{H}_0 + \lambda u(\xi) \right] \phi_k(\xi) &= \epsilon_k\phi_k(\xi) \quad.
\end{align}
Transforming the eigenstates into the basis of unperturbed eigenstates and expanding the eigenenergies and coefficients in Taylor series in $\lambda$,
\begin{align}
	\phi_k &= \sum_nc_{kn}\psi_n \quad,\quad \epsilon_k = \sum_p\lambda^p\epsilon_k^{(p)} \quad,\quad c_{km} = \sum_p\lambda^pc_{km}^{(p)} \quad,\quad u_{mn} = \int\limits_{-\infty}^\infty\psi_m^\ast(\xi)u(\xi)\psi_n(\xi)\,\text{d}\xi \quad,
\end{align}
yields
\begin{align}
	&\left(\epsilon_m^{(0)}-\epsilon_k^{(0)}\right)c_{km}^{(0)} + \sum_{p=1}^\infty \left[ \vphantom{\sum_{q=1}^{p-1}} \left(\epsilon_m^{(0)}-\epsilon_k^{(0)}\right)c_{km}^{(p)} - \epsilon_k^{(p)}\delta_{km} - \sum_{q=1}^{p-1}\epsilon_k^{(q)}c_{km}^{(p-q)} + \sum_nc_{kn}^{(p-1)}u_{mn} \right]\lambda^p = 0 \quad.
\end{align}
This gives recursion equations for the $p$th order corrections:
\begin{align}
	\epsilon_k^{(p)} &= \sum_{n\neq k}u_{kn}c_{kn}^{(p-1)} - \sum_{q=2}^{p-1}\epsilon_k^{(q)}c_{kk}^{(p-q)} \\
	c_{km}^{(p)} &= \frac{1}{\epsilon_k^{(0)}-\epsilon_m^{(0)}}\left( \sum_nu_{mn}c_{kn}^{(p-1)} - u_{kk}c_{km}^{(p-1)} - \sum_{q=2}^{p-1}\epsilon_k^{(q)}c_{km}^{(p-q)} \right) \quad.
\end{align}
Especially for the 1st, 2nd, and 3rd order corrections, one gets
\begin{align}
	\epsilon_k^{(1)} &= u_{kk} \quad,\\
	\epsilon_k^{(2)} &= -\sum_{n\neq k}\frac{u_{nk}u_{kn}}{\epsilon_n^{(0)}-\epsilon_k^{(0)}} \quad,\\
	\epsilon_k^{(3)} &= -\sum_{n\neq k}\frac{u_{nk}u_{kk}u_{kn}}{\left(\epsilon_n^{(0)}-\epsilon_k^{(0)}\right)^2} + \sum_{n\neq k}\sum_{l\neq k}\frac{u_{kn}u_{nl}u_{lk}}{\left(\epsilon_n^{(0)}-\epsilon_k^{(0)}\right)\left(\epsilon_l^{(0)}-\epsilon_k^{(0)}\right)} \quad.
\end{align}

\noindent Expanding $u(\xi)$ in a Taylor series in $\xi$,
\begin{align}
	u(\xi) &= \sum_j u_j\xi^j \quad,
\end{align}
gives
\begin{align}
	u_{mn} &= \sum_j u_j u_{mn}^{(j)} \quad,\\
	u_{mn}^{(j)} &= \frac{1}{\sqrt{2^{m+n}m!n!}\sqrt{\pi}} \int\limits_{-\infty}^\infty \xi^j\text{e}^{-\xi^2}H_m(\xi)H_n(\xi) \,\text{d}\xi \quad.
\end{align}
Inserting~\eqn{Hermite:recursive} gives the recursion
\begin{align}
	u_{mn}^{(j)} &= \sqrt{\frac{n+1}{2}}u_{m,n+1}^{(j-1)} + \sqrt{\frac{n}{2}}u_{m,n-1}^{(j-1)} \quad.
\end{align}

\begin{figure}[tbp]
	\includegraphics{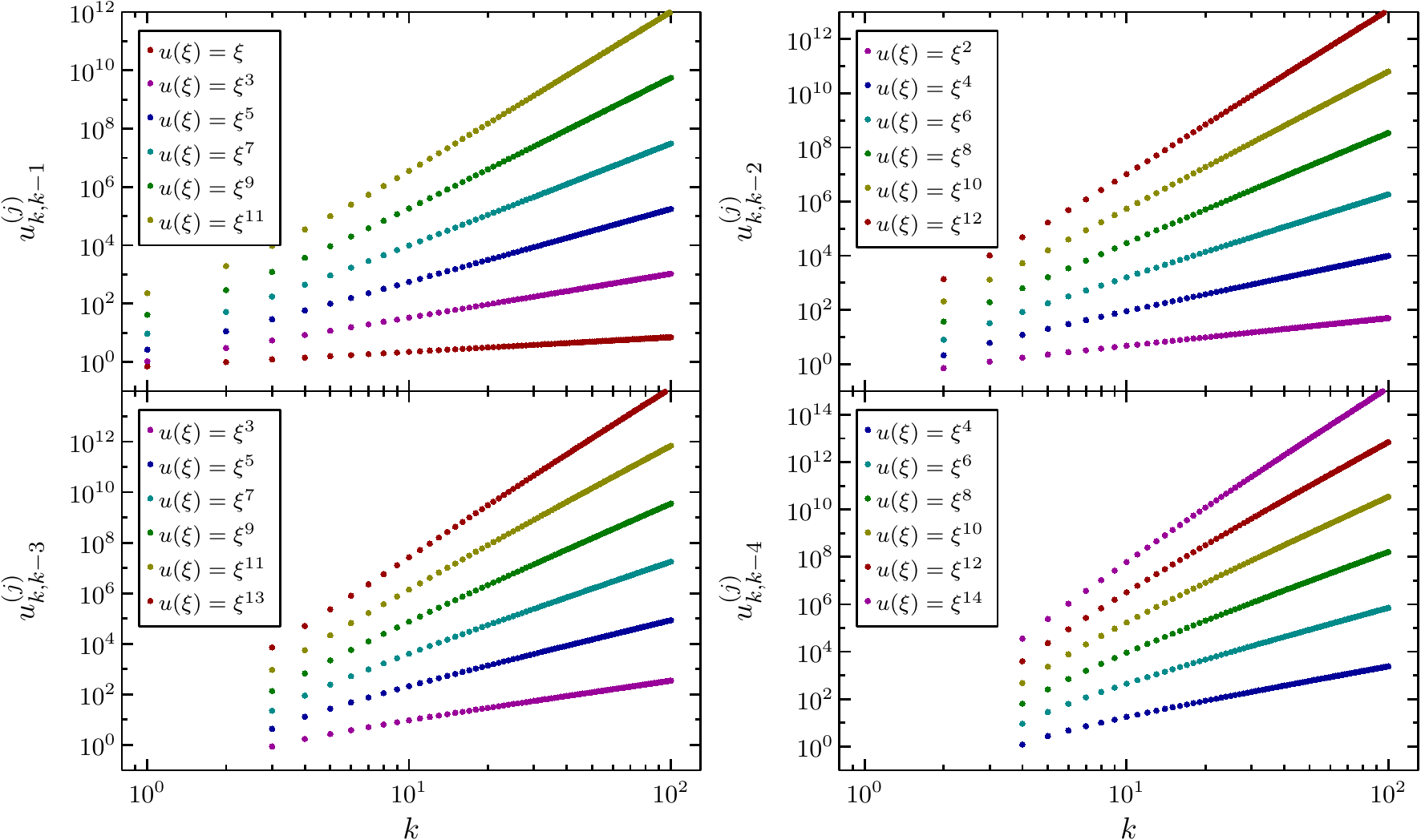}
	\caption{$u_{k,k-l}^{(j)}$ for different $l$ and for different perturbation potentials $u(\xi)=\xi^j$.}
	\label{fig:u_mn_j}
\end{figure}

\noindent The diagonal elements $u_{kk}^{(j)}$ are polynomials in $k$ of order $j/2$.
For the smallest $j$ one gets
\begin{align}
	u_{kk}^{(0)} &= 1 \quad,\\
	u_{kk}^{(2)} &= \frac{1}{2} + k \quad,\\
	u_{kk}^{(4)} &= \frac{3}{4} + \frac{3}{2}k + \frac{3}{2}k^2 \quad,\\
	u_{kk}^{(6)} &= \frac{15}{8} + 5k + \frac{15}{4}k^2 + \frac{5}{2}k^3 \quad,\\
	u_{kk}^{(8)} &= \frac{105}{16} + \frac{35}{2}k + \frac{175}{8}k^2 + \frac{35}{4}k^3 + \frac{35}{8}k^4 \quad.
\end{align}
The nondiagonal elements can be written as
\begin{align}
	u_{k,k-l}^{(j)} = \sqrt{\frac{k!}{2^l(k-l)!}}p_{kl}^{(j)} \quad ,
\end{align}
where $p_{kl}^{(j)}$ is a polynomial in $k$ of order $(j-l)/2$.
Thus, $(u_{k,k-l}^{(j)})^2$ is a polynomial in $k$ of order $j$ for all $l$.
For the smallest $j$  and $l$ one gets
\begin{align}
	p_{k,1}^{(1)} &= 1 \quad,\\
	p_{k,1}^{(3)} &= \frac{3}{2}k \quad,\\
	p_{k,1}^{(5)} &= \frac{5}{4}(1+2k^2) \quad,\\
	p_{k,1}^{(7)} &= \frac{35}{8}k(2+k^2) \quad,\\
	p_{k,2}^{(2)} &= 1 \quad,\\
	p_{k,2}^{(4)} &= -1+2k \quad,\\
	p_{k,2}^{(6)} &= \frac{15}{4}(1-k+k^2) \quad,\\
	p_{k,2}^{(8)} &= \frac{7}{2}(-1+2k)(3-k+k^2) \quad,\\
	p_{k,3}^{(3)} &= 1 \quad,\\
	p_{k,3}^{(5)} &= \frac{5}{2}(-1+k) \quad,\\
	p_{k,3}^{(7)} &= \frac{21}{4}(2-2k+k^2) \quad,\\
	p_{k,3}^{(9)} &= \frac{21}{4}(-1+k)(9-4k+2k^2) \quad,\\
	p_{k,4}^{(4)} &= 1 \quad,\\
	p_{k,4}^{(6)} &= \frac{3}{2}(-3+2k) \quad,\\
	p_{k,4}^{(8)} &= \frac{7}{2}(7-6k+2k^2) \quad,\\
	p_{k,4}^{(10)} &= \frac{15}{4}(-3+2k)(13-6k+2k^2) \quad.
\end{align}
The results are shown in \fig{u_mn_j}.

